\documentclass[journal=jctcce,manuscript=article,articletitle=true,layout=twocolumn]{achemso} 
\setkeys{acs}{articletitle=true}
\setkeys{acs}{doi = false} 
\setkeys{acs}{maxauthors=10}
\setkeys{acs}{etalmode=truncate}

\usepackage[version=4]{mhchem} 
\usepackage{amssymb,amsmath}
\usepackage{tabularx}
\usepackage[center]{subfigure}
\usepackage{multirow}
\usepackage{color}
\usepackage{comment}
\usepackage[usenames, dvipsnames]{xcolor}
\usepackage{graphicx}
\usepackage[colorlinks=true, allcolors=blue]{hyperref}
\usepackage{cleveref}
\usepackage{threeparttable}
\usepackage{attachfile}
\usepackage{caption}

\newcolumntype{C}{>{\centering\arraybackslash}X}
\newcolumntype{R}{>{\raggedleft\arraybackslash}X}
\newcolumntype{L}{>{\raggedright\arraybackslash}X}

\newcommand*\citeref[1]{ref.~\citenum{#1}}
\newcommand*\citerefs[1]{refs.~\citenum{#1}}
\newcommand*\jouletesla[1]{\ensuremath{#1 \times 10^{-30}\, {\rm J}/{\rm T}^2}}

\newcommand*\Gimic{\textsc{Gimic}}
\newcommand*\Numgrid{\textsc{Numgrid}}

\newcommand*\Gaussian{\textsc{Gaussian}}
\newcommand*\Turbomole{\textsc{Turbomole}}

\SectionNumbersOn

\author{Caspar J. Schattenberg}
\affiliation{Technische Universit\"at Berlin, Institut f\"ur Chemie,
Theoretische Chemie/Quantenchemie, Sekr. C7, Stra{\ss}e des 17. Juni
135, D-10623, Berlin, Germany}

\author{Artur Wody\'nski}
\affiliation{Technische Universit\"at Berlin, Institut f\"ur Chemie, Theoretische Chemie/Quantenchemie, Sekr. C7, Stra{\ss}e des 17. Juni 135, D-10623, Berlin, Germany}

\author{Hugo {\AA}str{\"o}m}
\affiliation{University of Helsinki, Department of Chemistry,
Faculty of Science, P.O. Box
  55 (A.I. Virtanens plats 1), FI-00014 University of Helsinki, Finland}

\author{Dage Sundholm}
\affiliation{University of Helsinki, Department of Chemistry,
Faculty of Science, P.O. Box
  55 (A.I. Virtanens plats 1), FI-00014 University of Helsinki, Finland}

\author{Martin Kaupp}
\email{martin.kaupp@tu-berlin.de}
\affiliation{Technische Universit\"at Berlin, Institut f\"ur Chemie,
Theoretische Chemie/Quantenchemie, Sekr. C7, Stra{\ss}e des 17. Juni
135, D-10623, Berlin, Germany}

\author{Susi Lehtola}
\email{susi.lehtola@alumni.helsinki.fi}
\affiliation{University of Helsinki, Department of Chemistry,
Faculty of Science, P.O. Box
  55 (A.I. Virtanens plats 1), FI-00014 University of Helsinki, Finland}
\alsoaffiliation{Molecular Sciences Software
  Institute, Blacksburg, Virginia 24061, United States}

\title{Revisiting Gauge-Independent Kinetic Energy Densities in Meta-GGAs and Local Hybrid Calculations of Magnetizabilities}

\abbreviations{NMR,GIMIC}

\begin{document}

\begin{tocentry}
  \centering
  \includegraphics[width=3.25in,height=1.75in,keepaspectratio]{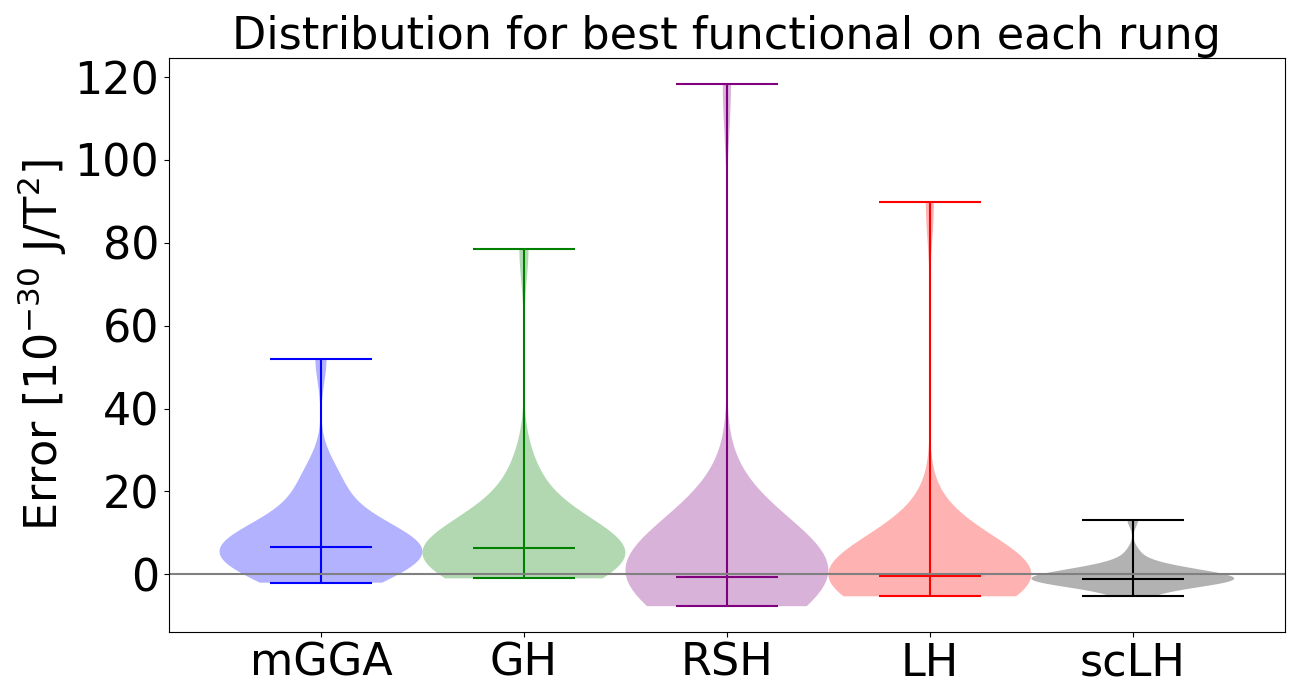}
\end{tocentry}


\begin{abstract}

In a recent study [J. Chem. Theory Comput. 2021, 17, 1457--1468], some
of us examined the accuracy of magnetizabilities calculated with
density functionals representing the local density approximation
(LDA), generalized gradient approximation (GGA), meta-GGA (mGGA) as
well as global hybrid (GH) and range-separated (RS) hybrid functionals
by assessment against accurate reference values obtained with
coupled-cluster theory with singles, doubles and perturbative triples
[CCSD(T)].  Our study was later extended to local-hybrid (LH)
functionals by Holzer et al.  [J. Chem. Theory Comput. 2021, 17,
  2928--2947]; in this work, we examine a larger selection of LH
functionals, also including range-separated LH (RSLH) functionals and
strong-correlation LH (scLH) functionals. Holzer et al also studied
the importance of the physically correct handling of the magnetic
gauge dependence of the kinetic energy density $(\tau)$ in mGGA
calculations by comparing the Maximoff--Scuseria formulation of $\tau$
used in our aforementioned study to the more physical current-density
extension derived by Dobson. In this work, we also revisit this
comparison with a larger selection of mGGA functionals. We find that
the newly tested LH, RSLH and scLH functionals outperform all the
functionals considered in the previous studies. The various LH
functionals afford the seven lowest mean absolute errors, while also
showing remarkably small standard deviations and mean errors. Most
strikingly, the best two functionals are scLHs that also perform
remarkably well in cases with significant multiconfigurational
character such as the ozone molecule, which is traditionally excluded
from the statistical error evaluation due to its large errors with
common density functionals.
\end{abstract}

\section{Introduction}
\label{sec:intro}

Molecular magnetic properties such as nuclear magnetic resonance (NMR)
shielding constants, NMR spin-spin coupling constants, and
magnetizabilities are useful probes of molecular and electronic
structure, and they can be studied to pinpoint the location of atoms
in a molecule, for instance. However, interpreting the measured
spectra often requires computational modeling.  Quantum chemical
modeling of magnetic properties is often pursued using Kohn--Sham (KS)
density functional theory\cite{Hohenberg1964_PR_864, Kohn1965_PR_1133}
(DFT).\cite{Helgaker1999_CR_293, Gauss2002_ACP_355, Keal2004_CPL_374,
  Gauss2007_JCP_74101, Lutnes2009_JCP_144104, Teale2013_JCP_24111,
  Loibl2014_JCP_24108, Reimann2017_JCTC_4089} However, the accuracy of
DFT for some of these properties has not been thoroughly established
in the literature. Some of us recently employed the benchmark set of
\citeref{Lutnes2009_JCP_144104} to assess the accuracy of density
functionals\cite{Lehtola2021_JCTC_1457, Lehtola2021_JCTC_4629} with an
emphasis on newer functionals from rungs 2--4 of the usual Jacob's
ladder hierarchy;\cite{Perdew2001_ACP_1} this benchmark set consists
of magnetizabilities for 28 small main-group molecules computed with
coupled-cluster theory with singles, doubles and perturbative triples
[CCSD(T)], which is a highly accurate wave function theory (WFT).

It was found in \citeref{Lehtola2021_JCTC_1457} that the
BHandHLYP\cite{Becke1993_JCP_1372} global hybrid (GH) functional and
some range-separated hybrid (RSH) functionals provided the closest
agreement with the CCSD(T) reference data, with the smallest mean
absolute deviations (MADs) being slightly above \jouletesla{3}, while
Hartree--Fock gave \jouletesla{7.22} and was ranked 29$^\text{th}$
best out of 52 methods evaluated.\cite{Lehtola2021_JCTC_1457,
  Lehtola2021_JCTC_4629} Some functionals, in particular of the
highly parameterized Minnesota functionals, were found to reach MADs
with large errors above \jouletesla{10}.

Our study in \citerefs{Lehtola2021_JCTC_1457} and
\citenum{Lehtola2021_JCTC_4629} employed
\Turbomole{},\cite{TBM_Balasubramani2020_JCP_184107, TBM} which until
recently relied on the Maximoff--Scuseria (MS)
approach\cite{Maximoff2004_CPL_408} to turn the kinetic energy density
$\tau$ used in meta-generalized gradient approximation (mGGA)
functionals into a gauge-independent quantity in the presence of a
magnetic field. However, recent work by some of us has shown that the
MS approach that is also used in the \Gaussian{}\cite{gaussian16}
program may lead to artefacts for nuclear magnetic shielding
constants, such as artificial paramagnetic contributions to the
shielding constant of spherical atoms, which can be avoided by
employing Dobson's current-density extension of
$\tau$\cite{Dobson1993_JCP_8870},
instead;\cite{Schattenberg2021_JCTC_1469, Schattenberg2021_JPCA_2697,
  Schattenberg2021_JCTC_7602, Schattenberg2021_JCTC_273} this approach
is nowadays available in \Turbomole{} for time-dependent density
functional theory (TDDFT) calculations\cite{Bates2012_JCP_164105,
  Bates2022_JCP_159902, Grotjahn2022_JCP_111102} and other
properties.\cite{Holzer2021_JCTC_2928, Franzke2022_JCP_31102} The
approach is also used in other packages as well for TDDFT and NMR
calculations.\cite{Liang2022_JCTC_3460, Neese2022_WCMS_1606} We also
note that Dobson's current-density extension for $\tau$ has been found
to be important in the context of calculations in explicit magnetic
fields.\cite{Tellgren2014_JCP_34101, Furness2015_JCTC_4169,
  Reimann2015_PCCP_18834, Irons2020_JPCA_1321, Irons2021_JCTC_2166}

\citet{Holzer2021_JCTC_2928} recently employed the methodology of
\citerefs{Lehtola2021_JCTC_1457} and \citenum{Lehtola2021_JCTC_4629}
to compare the effect of the MS and Dobson formulations of $\tau$ on
the magnetizabilities of two test sets: the aforementioned
theory-based test set of \citeref{Lutnes2009_JCP_144104}, and a test
set based on experimental data. Their results appeared at first sight
somewhat inconclusive, as the two test sets resulted in different
rankings of various functionals.

Several local hybrid (LH) functionals that employ a position-dependent
exact-exchange (EXX) admixture\cite{Jaramillo2003_JCP_1068,
  Maier2019_WIRCMS_1378} and that have been shown to exhibit promising
accuracy for nuclear magnetic shielding
constants\cite{Schattenberg2021_JCTC_1469, Schattenberg2021_JPCA_2697,
  Schattenberg2021_JCTC_7602, Schattenberg2021_JCTC_273} performed
excellently for the theory-based test set, with relative mean absolute
deviations in magnetizabilities below 1 \%. However,
\citet{Holzer2021_JCTC_2928} put more emphasis on the comparison to
experimental data, in which deviations for most functionals including
LH functionals were generally above 5 \% and were less systematic than
in the theory-based benchmark.

We argue that the comparison to these experimental data
does not afford a reliable assessment of the accuracy of density
functionals, as many of these data exhibit large error bars.
As LH functionals are of interest for many kinds of
properties, we revisit their performance for magnetizabilities in this
work, based on the theoretical dataset of
\citet{Lutnes2009_JCP_144104} which was also used in previous
studies.\cite{Lehtola2021_JCTC_1457, Lehtola2021_JCTC_4629,
  Holzer2021_JCTC_2928} Although the limitations of this kind of
static DFT benchmarks need to be
recognized,\cite{Weymuth2022_PCCP_14692} such studies often give
helpful guidance in terms of the suitable physical contents of density
functional approximations.

Going beyond the LH functionals studied in
\citeref{Holzer2021_JCTC_2928}, we also include the first modern
range-separated local hybrid (RSLH)
$\omega$LH22t\cite{Fuerst2023_JCTC_488} in our assessment, as it has
been shown to provide remarkable accuracy for quasiparticle energies
of a wide variety of organic chromophores of interest in molecular
electronics and organic photovoltaics,\cite{Fuerst2023_JCTC_sub} while
also performing well for many other ground- and excited-state
properties.\cite{Fuerst2023_JCTC_488} We also investigate the
performance of recent strong-correlation-corrected LH functionals
(scLH) such as scLH22t and scLH22ta,\cite{Wodynski2022_JCTC_6111} as
well as the most recent models with simplified constructions of the
sc-correction terms.\cite{Wodynski2023_JCP_244117} We will show that
these functionals can more reliably reproduce magnetizabilities of
systems with large static correlation effects such as \ce{O3}, whose
magnetizabilities predicted by regular LH functionals significantly
deviate from the CCSD(T) reference value.\cite{Holzer2021_JCTC_2928}

In addition, we will also study the Dobson formulation for $\tau$ on a
wider set of $\tau$-dependent functionals than those studied in
\citeref{Holzer2021_JCTC_2928}, including various older and newer mGGA
functionals and mGGA-based global and range-separated hybrid
functionals.

The layout of this work is as follows. Next, in \cref{sec:theory}, we
will discuss the employed methodology for computing the
magnetizability (\cref{sec:magmethods}), gauge origin problems
(\cref{sec:gaugeproblems}), and LH functionals
(\cref{sec:local-hybrids}), and then the employed computational
methodology in \cref{sec:methods}. We present the results of this
study in \cref{sec:results}, and conclude in \cref{sec:conclusions}.
Atomic units are used throughout, unless specified otherwise.

\section{Theory}
\label{sec:theory}

\subsection{Methods for Calculating Magnetizabilities}
\label{sec:magmethods}

Magnetizabilities are commonly calculated as the second derivative of
the electronic energy with respect to the external magnetic
field\cite{Ruud1993_JCP_3847, Ruud1994_JACS_10135, Ruud1995_CP_157,
  Loibl2014_JCP_24108, Helgaker2012_CR_543}
\begin{equation} \xi_{\alpha\beta} = -\frac{\partial^2 E}{\partial
B_{\alpha} \partial  B_{\beta}}\Bigg|_{{\bf B}={\bf 0}}.
\label{eq:zeta-secondder} \end{equation}
The magnetic interaction can also be expressed as an integral over the
magnetic interaction energy density $\rho^{\bf B}({\bf r})$, which is
the scalar product of the magnetically induced current density
$\mathbf{J^B}(\mathbf{r})$ with the vector potential
$\mathbf{A^B}(\mathbf{r})$ of the external magnetic field
$\mathbf{B}$\cite{Stevens1963_JCP_550, Jameson1979_JPC_3366,
  Jameson1980_JCP_5684, Fowler1998_JPCA_7297, Ilias2013_MP_1373,
  Lazzeretti2000_PNMRS_1, Lazzeretti2018_JCP_134109,
  Lehtola2021_JCTC_1457, Lehtola2021_JCTC_4629}
\begin{equation}
E = \int \rho^{\bf B}({\bf r}) ~\mathrm{d}^3 r =
-\frac{1}{2}\int \mathbf{A}^{\mathbf{B}}(\mathbf{r}) \cdot
\mathbf{J^B}(\mathbf{r}) ~\mathrm{d}^3r.
\label{eq:Emag}
\end{equation}
The second derivatives of the magnetic interaction energy with respect
to the components of the external magnetic field, together forming the
elements of the magnetizability tensor $\boldsymbol{\xi}$, can be
obtained from \cref{eq:Emag} as an integral over the scalar product of
the first derivatives of the vector potential of the external magnetic
field with the magnetically induced current-density susceptibility
(CDT), $ \partial J_\gamma^{\bf B}(\mathbf{r}) / {\partial B_\beta}
\big|_{{\bf B}={\bf 0}}$, in the limit of a vanishing magnetic
field\cite{Sambe1973_JCP_555, Lazzeretti2000_PNMRS_1,
  Lazzeretti2018_JCP_134109}
\begin{equation}
  \xi_{\alpha\beta} =
\frac{1}{2}\int \frac{\partial \mathbf{A}^{\mathbf{B}}(\mathbf{r})}{\partial B_\alpha}  \cdot
\frac{\partial \mathbf{J^B}(\mathbf{r})}{\partial B_\beta}  ~\mathrm{d}^3r
\Bigg|_{{\bf B}={\bf 0}},
  \label{eq:magnetizability}
\end{equation}
where the vector potential $\mathbf{A}^\mathbf{B}(\mathbf{r})$ of a
homogeneous external magnetic field is
\begin{equation}
  \mathbf{A}^\mathbf{B}(\mathbf{r}) = \frac{1}{2} \mathbf{B} \times (\mathbf{r}-\mathbf{R}_O)
  \label{eq:vecpot}
\end{equation}
and $\mathbf{R}_O$ is an arbitrary gauge origin.
The CDT can be calculated in a given atomic-orbital
(AO) basis set from the unperturbed and the magnetically perturbed AO
density matrices, which can be obtained at any level of theory from
nuclear magnetic shielding calculations, for
example.\cite{Juselius2004_JCP_3952, Taubert2011_JCP_54123,
  Fliegl2011_PCCP_20500, Sundholm2016_WIRCMS_639}

The isotropic magnetizability ($\overline{\xi}$) is obtained as one
third of the trace of the magnetizability tensor
\begin{equation}
  \overline{\xi}=\frac{1}{3} \mathrm{Tr}~\boldsymbol{\xi} =
  \int \overline{\rho}^\xi({\bf r}) \mathrm{d}^3 r ,
  \label{eq:xiave}
\end{equation}
where the magnetizability density tensor is defined in terms of the
CDT as
\begin{equation}
\rho_{\alpha\beta}^\xi ({\bf r}) = \frac{1}{2} \sum_{\delta \gamma}
\epsilon_{\alpha\delta\gamma} {r}_\delta
\frac {\partial J_\gamma^{\bf B}(\mathbf{r})} {\partial B_\beta}
\Bigg|_{{\bf B}={\bf 0}} ,
\label{eq:magdens}
\end{equation}
where $\epsilon_{\alpha \delta \gamma}$ is the Levi--Civita symbol,
and $\alpha$, $\beta$, $\gamma$, $\delta$ and $r_\delta$ $\in
\{x,y,z\}$ are Cartesian directions.  The $\xi_{\alpha \alpha}, \alpha
\in \{x,y,z\}$ elements of the magnetizability tensor can be obtained
by quadrature as
\begin{equation}
\xi_{\alpha \alpha} = \sum_{i = 1}^{n} w_i \rho^\xi_{i;\alpha \alpha}
\label{eq:quadrature}
\end{equation}
where $\rho^\xi_{i;\alpha \alpha}$ is a diagonal element of the
magnetizability density tensor at quadrature point $i$ and $w_i$ is
the corresponding quadrature weight.  

The use of the above quadrature scheme allows magnetizabilities to be
computed even when the analytical second derivatives required to
evaluate \cref{eq:zeta-secondder} are not
available.\cite{Lehtola2021_JCTC_1457} In addition to providing a
complementary approach to computing magnetizabilities, the numerical
integration approach also affords information about spatial
contributions to the magnetizability;\cite{Lehtola2021_JCTC_1457,
  Lehtola2021_JCTC_4629} similar approaches have previously also been
used to study spatial contributions to nuclear magnetic shielding
constants.\cite{Steiner2004_PCCP_261, Pelloni2004_OL_4451,
  Ferraro2004_CPL_268, Soncini2005_CPL_164, Ferraro2005_MRC_316,
  Acke2018_JCC_511, Acke2019_PCCP_3145, Jinger2021_JPCA_1778,
  Fliegl2021_C_1005, Summa2021_CCA_43}

\subsection{Gauge-Origin Problems}
\label{sec:gaugeproblems}

The use of a finite one-particle basis set introduces issues with
gauge dependence into quantum chemical calculations of magnetic
properties. The CDT can be made gauge-origin
independent\cite{Juselius2004_JCP_3952, Fliegl2011_PCCP_20500,
  Sundholm2016_WIRCMS_639, Sundholm2021_CC_12362} by using
gauge-including atomic orbitals (GIAOs),\cite{Hameka1962_RMP_87,
  Ditchfield1974_MP_789, Wolinski1990_JACS_8251, Gauss1992_CPL_614,
  Gauss2002_ACP_355} also known as London atomic orbitals
(LAOs),\cite{London1937_JPlR_397, Ruud1993_JCP_3847,
  Helgaker2012_CR_543}
\begin{equation} \chi_\mu(\mathbf{r}) = e^{-i (\mathbf{B}\times
[\mathbf{R}_\mu-\mathbf{R}_O] \cdot \mathbf{r})/(2c)} \chi_\mu^{(0)}
\left(\mathbf{r}\right), \label{eq:lao} \end{equation}
where $i$ is the imaginary unit, $\chi_\mu^{(0)}(\mathbf{r})$ is a
basis function centered at $\mathbf{R}_\mu$, and $c$ is the speed of
light which has the value $c=\alpha^{-1} \approx 137.036$ in atomic units,
where $\alpha$ is the fine-structure constant.

The mGGA approximation for the exchange-correlation energy density
contains a dependence on the kinetic energy density $\tau$, which
reads in the field-free case as
\begin{equation}
\tau(\mathbf{r}) = \frac 1 2 \sum_{\mu\nu} D_{\mu\nu}
\nabla\chi_\mu^*(\mathbf{r}) \cdot \nabla\chi_\nu^{ }(\mathbf{r}),
\label{eq:KStau}
\end{equation}
where $D_{\mu\nu}$ is the AO density matrix. However, this dependence
requires additional care, as the (uncorrected) $\tau(\mathbf{r})$ of
\cref{eq:KStau} is clearly not \emph{a priori} gauge invariant even
when using GIAOs.

A widely used model to render $\tau(\mathbf{r})$ gauge-invariant was
introduced by \citet{Maximoff2004_CPL_408} (MS) as
\begin{equation}
\label{eq:tau-MS}
\tau_\text{MS}^{ }(\mathbf{r}) = \tau(\mathbf{r}) +
\frac{1}{c} (\mathbf{A}^\mathbf{B}) \cdot \mathbf{j}_p(\mathbf{r}) +
\frac{|\mathbf{A}^\mathbf{B}|^2}{c^2} \rho(\mathbf{r}),
\end{equation}
where $\mathbf{j}_p^{ }(\mathbf{r})$ is the paramagnetic current density
defined as
\begin{equation}
\label{eq:paramagnetic}
\mathbf{j}_p^{ }(\mathbf{r}) = \frac{i}{2} \sum_{\mu\nu} D_{\mu\nu}^{ } (
\chi_\nu^{ } \nabla\chi_\mu^* - \chi^*_\mu \nabla\chi_\nu^{ } ).
\end{equation}

Advantages for coupled-perturbed KS (CPKS) calculations with mGGAs
arise from the diagonality of the Hessian for $\tau_{MS}^{
}(\mathbf{r})$, which allows solving the CPKS equations in a single
step. However, the semi-local exchange-correlation (XC) contribution
does not produce a linear response, even though a genuine
current-density functional is expected to provide such a
response.\cite{Vignale1987_PRL_2360,Vignale1988_PRB_10685}
$\tau_{MS}^{ }(\mathbf{r})$ also does not constitute a proper iso-orbital
indicator,\cite{Sagvolden2013_MP_1295} is
non-universal,\cite{Tao2005_PRB_205107,Bates2012_JCP_164105,Reimann2015_PCCP_18834}
and introduces paramagnetic artefacts in shielding
calculations.\cite{Schattenberg2021_JCTC_1469}

A formulation that avoids the disadvantages of the MS model was
proposed by Dobson\cite{Dobson1992_JPCM_7877, Dobson1993_JCP_8870}
\begin{equation}
\label{eq:tau-D}
\tau_{\rm D}^{ }(\mathbf{r}) = \tau(\mathbf{r}) -\frac{|\mathbf{j}_p^{ }(\mathbf{r})|^2}{2\rho(\mathbf{r})}.
\end{equation}
$\tau_\text{D}(\mathbf{r})$ in \cref{eq:tau-D} leads to gauge independence,
while also introducing a current-response of the semi-local XC
contribution and thereby a correct physical behavior for $\tau$-dependent
density functionals. As the electronic Hessian corresponding to
\cref{eq:tau-D} is non-diagonal, an iterative solution to the
CPKS equations is now necessary, rendering the calculations slightly
more expensive than when the physically incorrect \cref{eq:tau-MS} is used.

\subsection{Local Hybrid Functionals}
\label{sec:local-hybrids}

In addition to studying the importance of the Dobson formulation
(\cref{eq:tau-D}) of the kinetic-energy density in reproducing
accurate magnetizabilities, we will also evaluate the accuracy of LH
functionals. A form for an LH functional that emphasizes the inclusion
of nonlocal correlation terms (often considered to cover nondynamical
correlation, NDC) together with full exact exchange and a (semi-local)
dynamical correlation (DC) functional is\cite{Maier2019_WIRCMS_1378,
  Perdew2001_ACP_1}
\begin{align}
  \label{eq:ndc-lh}
	E_{\text{XC}}^{\text{LH}} &= E_\text{X}^{ex}  + \int  \left( 1-g(\mathbf{r}) \right) \times\nonumber\\
				&\:\:  \left( e_{\text{X}}^{\text{sl}}(\mathbf{r}) +G(\mathbf{r}) -e_{\text{X}}^{\text{ex}}(\mathbf{r}) \right) \text{d}\mathbf{r} + E_\text{C}^{ } \nonumber \\
					\phantom{\sum}&=E_\text{X}^{\text{ex}} + E_{\text{NDC}}^{ } + E_{\text{DC}}^{ },
\end{align}
where $g(\mathbf{r})$ is the local mixing function (LMF) controlling
the fraction of exact exchange included at $\mathbf{r}$. In most of
the LH functionals considered here we use a so-called t-LMF defined as
the scaled ratio between the von Weizsäcker\cite{Weizsacker1935_431}
and KS kinetic energy densities
%
\begin{align}
	g(\mathbf{r})=a\cdot\frac{\tau_{\text{W}}(\mathbf{r})}{\tau(\mathbf{r})}, \qquad\tau_{\text{W}}(\mathbf{r}) = \frac{|\nabla\rho(\mathbf{r})|^2}{8\rho(\mathbf{r})}.
\end{align}

The calibration function (CF) $G(\mathbf{r})$ is used in
\cref{eq:ndc-lh} to correct for the ambiguity of the semi-local and
exact exchange-energy densities.\cite{Burke1998_JCP_8161,
  Cruz1998_JPCA_4911, Maier2016_PCCP_21133, Maier2019_WIRCMS_1378,
  Tao2008_PRA_012509}
In the LH functionals with a CF from the Berlin
group,\cite{Haasler2020_JCTC_5645, Fuerst2023_JCTC_488,
  Wodynski2022_JCTC_6111} the semi-local CFs are currently derived
within the partial integration gauge (pig)
approach.\cite{Maier2016_PCCP_21133}

We also evaluate two recent extensions of LH functionals: the
so-called strong-correlation-corrected LH functionals
(scLH)\cite{Wodynski2021_JCP_144101, Wodynski2022_JCTC_6111,
  Wodynski2023_JCP_244117} and a range-separated local hybrid (RSLH)
functional ($\omega$LH22t).\cite{Fuerst2023_JCTC_488} In the scLH
functionals, a strong-correlation factor $q_{\text{AC}}(\mathbf{r})$
is introduced into the LMF of the LH:
\begin{align}\label{eq:scLH} &E_{\text{XC}}^{\text{scLH}} =
E_\text{X}^{\text{ex}}  + 2 \int  q_{\text{AC}}^{ }(\mathbf{r})\cdot \bigg[
\left( 1-g(\mathbf{r}) \right) \times  \nonumber\\ \phantom{\sum}&\:\:  \left(
e_{\text{X}}^{\text{sl}}(\mathbf{r}) +G(\mathbf{r})
-e_{\text{X}}^{\text{ex}}(\mathbf{r}) \right) + e_\text{C}^{ }(\mathbf{r})
\bigg]\text{d}\mathbf{r}.  \end{align}
This approach is based on the local adiabatic connection approach, and
is adapted from the KP16/B13\cite{Kong2016_JCTC_133} and
B13\cite{Becke2013_JCP_74109} sc-models. The most recent models employ
simplified real-space measures to detect strong correlations, as well
as modified damping functions to avoid double-counting of NDC
contributions in more weakly correlated
situations.\cite{Wodynski2023_JCP_244117}

In the absence of strong correlations, $q_{\text{AC}}^{ }
\longrightarrow 0.5$ as a lower bound, and the underlying LH
functional is restored. Whenever the quantities underlying
$q_{\text{AC}}^{ }$ detect locally the presence of strong
correlations, $q_{\text{AC}}^{ }$ is increased, maximally up to
1.0. In the exchange picture, this means that the EXX admixture is
locally diminished; in some cases it may even become
negative.\cite{Wodynski2022_JCTC_6111, Wodynski2023_JCP_244117} This
enhances the simulation of nonlocal correlation contributions and is
crucial for reducing fractional spin errors and for improving
spin-restricted bond dissociation curves.\cite{Wodynski2022_JCTC_6111,
  Wodynski2023_JCP_244117}

RSLH functionals like $\omega$LH22t\cite{Fuerst2023_JCTC_488} may be
written as
\begin{align}
  \label{eq:rslh}
  E_{\text{XC}}^{\text{LH}} &= E_\text{X}^{ex}  + \int  \left( 1-g(\mathbf{r}) \right) \times\nonumber\\
  \phantom{\sum}&\:\:  \left( e_{\text{X,sr}}^{\text{sl}}(\mathbf{r}) +G(\mathbf{r}) -e_{\text{X,sr}}^{\text{ex}}(\mathbf{r}) \right) \text{d}\mathbf{r} + E_\text{C}^{ },
\end{align}
where $e_{\text{X,sr}}^{\text{sl}}$ and $e_{\text{X,sr}}^{\text{ex}}$
are short-range exchange-energy densities, controlled by the
range-separation parameter $\omega$. In consequence, the $\omega$LH22t
functional has full long-range EXX
admixture,\cite{Fuerst2023_JCTC_488} like the RSH functionals
evaluated here as well, while the short-range EXX admixture is determined by the LMF.

\section{Computational Methods}
\label{sec:methods}

As in \citerefs{Lehtola2021_JCTC_1457} and \citenum{Lehtola2021_JCTC_4629}, the
unperturbed and magnetically perturbed density matrices are generated with the
nuclear magnetic shielding module of the \Turbomole{} program
(\texttt{mpshift}).\cite{TBM} A locally modified version of \Turbomole{} 7.6 has been
used for the present calculations.
\Turbomole{} employs \textsc{Libxc}\cite{Lehtola2018_S_1} to evaluate
many of the presently considered density functionals. A detailed description
of the implementations of the LH
functionals,\cite{Schattenberg2021_JCTC_1469} the Dobson
model,\cite{Schattenberg2021_JCTC_273} and higher derivatives of the
density used in the pig2 CF\cite{Schattenberg2021_JPCA_2697} can be
found in the respective publications. The necessary equations for
the magnetic-field derivatives of scLHs and RSLHs are
outlined in Section S1 of the Supporting Information.

The calculations were carried out with the aug-cc-pCVQZ basis
set\cite{Dunning1989_JCP_1007, Kendall1992_JCP_6796,
  Woon1993_JCP_1358, Woon1995_JCP_4572, Peterson2002_JCP_10548,
  Prascher2011_TCA_69} (with aug-cc-pVQZ on hydrogen atoms) employing
GIAOs\cite{London1937_JPlR_397, Ditchfield1972_JCP_5688,
  Ditchfield1974_MP_789, Wolinski1990_JACS_8251}
a.k.a. LAOs\cite{Ruud1993_JCP_3847, Helgaker2012_CR_543}. The
self-consistent field convergence criteria were set as $10^{-9}$ for
the energy and $10^{-7}$ for the density. Large numerical integration
grids per the approach of \citet{Becke1988_JCP_2547} were used with
the \Turbomole{}\cite{TBM_Balasubramani2020_JCP_184107, TBM} setting
\texttt{gridsize 7}, following the original work of
\citet{Treutler1995_JCP_346} with later extensions documented in the
\Turbomole{} manual.\bibnote{\Turbomole{} User's Manual,
  \url{https://www.turbomole.org/downloads/doc/Turbomole_Manual_76.pdf},
  accessed 27 October 2023.}
The non-standard exact-exchange integrals occurring in LH, scLH and
RSLH functionals are calculated by efficient semi-numerical
integration techniques, \cite{Bahmann2015_JCTC_1540,
  Liu2017_JCTC_2571, Laqua2018_JCTC_3451, Holzer2020_JCP_184115} using
standard DFT grids.

The resolution-of-the-identity (RI) approximation was used to evaluate
the Coulomb contribution (J), using Turbomole's ``universal''
auxiliary basis set by \citeauthor{Weigend2006_PCCP_1057}.\cite{Weigend2006_PCCP_1057}  Although one
needs to be careful about mixing auxiliary basis sets for different
families, we tested the accuracy of this RI-J approximation with a few
of the functionals considered in this work, and the resulting
magnetizabilities with the RI-J approximation coincided with values
obtained without it to the reported number of digits, which is
consistent with results recently computed for NMR
shieldings.\cite{Stoychev2018_JCTC_619, Reiter2018_JCTC_191}
We note that accurate auxiliary basis sets with controllable accuracy
for RI calculations can nowadays be easily generated
automatically,\cite{Lehtola2021_JCTC_6886, Lehtola2023_JCTC_6242} and
recommend such autogenerated auxiliary basis sets to be used when
tailored basis sets are not available. We refer the reader to
\citeref{Pedersen2023_WIRCMS_1692} for a recent review of further
automatical auxiliary basis generation techniques.

The magnetizabilities are computed from the density matrices produced
by \Turbomole{} with the \Gimic{} program.\cite{Juselius2004_JCP_3952,
  Taubert2011_JCP_54123, Fliegl2011_PCCP_20500,
  Sundholm2016_WIRCMS_639} The integral in \cref{eq:magdens} is
calculated in \Gimic{} using Becke's\cite{Becke1988_JCP_2547}
multicenter quadrature scheme,\cite{Lehtola2021_JCTC_1457,
  Lehtola2021_JCTC_4629} employing the \Numgrid{}
library\cite{Bast:20} to generate the atomic quadrature grids with a
hardness parameter of 3 for the atomic weight partitioning, radial
grids from \citet{Lindh2001_TCA_178} and Lebedev's angular
grids.\cite{Lebedev1995_RASDM_283} Both \Gimic{} and \Numgrid{} are
free and open-source software.\cite{gimic-download, numgrid-download}

In this work, we study the accuracy of magnetizabilities reproduced by the 31
functionals listed in \cref{tab:ldas-ggas}
for a dataset of 28 molecules:
\ce{AlF}, \ce{C2H4}, \ce{C3H4}, \ce{CH2O}, \ce{CH3F}, \ce{CH4}, \ce{CO},
\ce{FCCH}, \ce{FCN}, \ce{H2C2O}, \ce{H2O}, \ce{H2S}, \ce{H4C2O}, \ce{HCN},
\ce{HCP}, \ce{HF}, \ce{HFCO}, \ce{HOF}, \ce{LiF}, \ce{LiH}, \ce{N2}, \ce{N2O},
\ce{NH3}, \ce{O3}, \ce{OCS}, \ce{OF2}, \ce{PN}, and \ce{SO2}, which have also
been used as benchmark molecules in other studies.\cite{Lutnes2009_JCP_144104,
Lehtola2021_JCTC_1457, Lehtola2021_JCTC_4629} The obtained DFT
magnetizabilities are compared to the CCSD(T) reference values of
\citeref{Lutnes2009_JCP_144104}. Since the magnetizability for \ce{O3}
introduces significant uncertainties due to large static correlation effects,
it was omitted from the statistical analysis, as was also done in
\citerefs{Lehtola2021_JCTC_1457} and \citenum{Lehtola2021_JCTC_4629}.

\begin{table*}
\footnotesize

\caption{The employed local-hybrid functionals (LH), range-separated
  LH functionals (RSLH), strong correlation LH functionals (scLH),
  functionals at the meta-generalized gradient approximation (mGGA),
  global hybrid (GH) functionals as well as
  range-separated (RS) GGA and mGGA functionals, and one GGA
  functional
      \label{tab:ldas-ggas}}
\begin{tabularx}{1.01\textwidth}{@{}llLr@{}}
      \hline
Functional       &  Type & Notes &  References \\
      \hline \hline
scLH23t-mBR$^{a}$		 & scLH	    & \mbox{ct-LMF ($a\approx 0.715$), damped $q_\text{AC-erf}^{ }$, pig2-CF, X$_{0.22\text{S}+0.78\text{PBE}}$+C$_\text{modB95}$} 	& \phantom{R}\citenum{Wodynski2023_JCP_244117} \\
scLH23t-mBR-P$^{a}$	 & scLH	    & \mbox{ct-LMF ($a\approx 0.715$), damped $q_\text{AC-Pad\'e}^{ }$, pig2-CF, X$_{0.22\text{S}+0.78\text{PBE}}$+C$_\text{modB95}$}  	& \phantom{R}\citenum{Wodynski2023_JCP_244117} \\
scLH22t			 & scLH		& \mbox{ct-LMF ($a=0.715$), damped $q_\text{AC}^{ }$, pig2-CF, X$_{0.22\text{S}+0.78\text{PBE}}$+C$_\text{modB95}$}  	& \phantom{R}\citenum{Wodynski2022_JCTC_6111} \\
scLH22ta		 & scLH	 	& \mbox{ct-LMF ($a=0.766$), $q_\text{AC}^{ }$, pig2-CF, X$_{0.04\text{S}+0.96\text{PBE}}$+C$_\text{modB95}$}  		& \phantom{R}\citenum{Wodynski2022_JCTC_6111} \\
scLH21ct-SVWN-m	 & scLH		& ct-LMF ($a=0.628$), $q_\text{AC}^{ }$, X$_\text{S}$+C$_\text{VWN}$  		& \phantom{R}\citenum{Wodynski2021_JCP_144101} \\
$\omega$LH22t	 & RSLH     & ct-LMF ($a=0.587$), $\omega=0.233$, pig2-CF, X$_{\text{PBE}}^{ }$+C$_\text{modB95}$  	& \phantom{R}\citenum{Fuerst2023_JCTC_488} \\
LH20t            & LH      	& ct-LMF ($a=0.715$), pig2-CF, X$_{0.22\text{S}+0.78\text{PBE}}$+C$_\text{modB95}$	& \phantom{R}\citenum{Haasler2020_JCTC_5645}        \\
LH20t nonCal     & LH      	& ct-LMF ($a=0.715$), X$_{0.22\text{S}+0.78\text{PBE}}$+C$_\text{modB95}$	& \phantom{R}\citenum{Haasler2020_JCTC_5645}              \\
LH14t-calPBE     & LH     	& \mbox{t-LMF ($a=0.5$), pig1-CF, X$_{0.49\text{S}+0.51\text{PBE}}$+C$_{0.55\text{PW92}+0.45\text{PBE}}$}    		& \phantom{R}\citenum{Arbuznikov2014_JCP_204101}             		\\
LH12ct-SsirPW92  & LH      	& ct-LMF ($a=0.646$), X$_\text{S}$+C$_\text{sicPW92}$ & \phantom{R}\citenum{Arbuznikov2012_JCP_14111}            		\\
LH12ct-SsifPW92  & LH      	& ct-LMF ($a=0.709$), X$_\text{S}$+C$_\text{sicPW92}$ & \phantom{R}\citenum{Arbuznikov2012_JCP_14111}       	        \\
LH07t-SVWN       & LH      	& t-LMF ($a=0.48$), X$_\text{S}$+C$_\text{VWN}$	   & \phantom{R}\citenum{Bahmann2007_JCP_11103,Kaupp2007_JCP_194102}	\\
LH07s-SVWN       & LH      	& s-LMF ($b=0.277$),  X$_\text{S}$+C$_\text{VWN}$ 	& \phantom{R}\citenum{Arbuznikov2007_CPL_160}             \\
mPSTS-a1$^{b}$         & LH      	& modified PSTS functional & \phantom{R}\citenum{Perdew2008_PRA_52513,Holzer2021_JCTC_2928}  \\
mPSTS-noa2$^{b}$      & LH      	& modified PSTS functional & \phantom{R}\citenum{Perdew2008_PRA_52513,Holzer2021_JCTC_2928}  \\
LHJ14            & LH      	& z-LMF ($c=0.096$), X$_\text{B88}$+C$_\text{B88}$    	& \phantom{R}\citenum{Johnson2014_JCP_124120}            \\
B97M-V           & mGGA   	&      & \phantom{R}\citenum{Mardirossian2015_074111}  \\
VSXC             & mGGA  	&      & \phantom{R}\citenum{VanVoorhis1998_JCP_400}               \\
MN15-L           & mGGA   	&      & \phantom{R}\citenum{Yu2016_1280}    \\
TPSS             & mGGA   	&      & \phantom{R}\citenum{Tao2003_146401, Perdew2004_6898} \\
M06-L            & mGGA   	&      & \phantom{R}\citenum{Zhao2006_194101}  \\
$\tau$-HCTH      & mGGA   	&      & \phantom{R}\citenum{Boese2002_JCP_9559}            \\
PW6B95           & GH, mGGA &      & \phantom{R}\citenum{Zhao2005_5656}   \\
TPSSh            & GH, mGGA &     	& \phantom{R}\citenum{Tao2003_146401, Perdew2004_6898} \\
M06-2X           & GH, mGGA &     	& \phantom{R}\citenum{Zhao2008_215}  \\
M06              & GH, mGGA &     	& \phantom{R}\citenum{Zhao2008_215}  \\
MN15             & GH, mGGA &     	& \phantom{R}\citenum{Yu2016_5032}   \\
B3LYP5$^{c}$            & GH, GGA  &     & \phantom{R}\citenum{Stephens1994_11623,Hertwig1997_CPL_345} \\
BHandHLYP        & GH, GGA &     	& \phantom{R}\citenum{Becke1993_1372} \\
$\omega$B97M-V   & RSH, mGGA &     	& \phantom{R}\citenum{Mardirossian2016_JCP_214110}             \\
$\omega$B97X-V   & RSH, GGA  &     & \phantom{R}\citenum{Mardirossian2014_9904}   \\
      \hline \hline
\multicolumn{4}{@{}p{17.5cm}@{}}{\scriptsize{$^{a}$ Use of simplified $q_{AC}^{ }$ based on error function (scLH23t-mBR) or Pad\'e functions (scLH23t-mBR-P); simplified construction of underlying function to identify regions of strong-correlation; see \citeref{Wodynski2023_JCP_244117} for details}} \\
\multicolumn{4}{@{}L}{\scriptsize{$^{b}$ See \citeref{Holzer2021_JCTC_2928} for further details of the a1 and noa2 models}} \\
\multicolumn{4}{@{}p{17.5cm}@{}}{\scriptsize{$^{c}$ This functional is called B3LYP in \Turbomole{}, and as discussed by \citeauthor{Hertwig1997_CPL_345},\cite{Hertwig1997_CPL_345} it is slightly different from the B3LYP functional used in \citeref{Lehtola2021_JCTC_1457}.}}
\end{tabularx}
\end{table*}

\section{Results}
\label{sec:results}

We will begin the discussion of the results in
\cref{sec:dobson-effect} by examining the accuracy of all the
considered functionals with the Dobson and MS models for $\tau$ in the
subset of data without \ce{O3}, which exhibits strong correlation
effects. We then discuss the accuracy of the various functionals
\ce{O3} in \cref{sec:O3}.  The magnetizabilities for all studied
molecules and functionals with the MS and Dobson formulations of
$\tau$ can be found in the Supporting Information (SI), accompanied
with violin plots of the corresponding error distributions.

\subsection{Accuracy of Density Functionals with Various Models for $\tau$}
\label{sec:dobson-effect}

\Cref{tab:maemestd} summarizes the overall statistical evaluations and
the ranking of the various functionals within the Dobson and
the MS models for $\tau$. We begin by noting that our
results agree with those of \citet{Holzer2021_JCTC_2928} for the
subset of functionals also studied in \citeref{Holzer2021_JCTC_2928}.

\begin{table*}
\begin{tabular}{llrrr|lrrr}
\hline
Rank & Functional      & MAE & ME & STD  & Rank & MAE & ME & STD \\
     & & \multicolumn{3}{c|}{$\tau_D^{}$} & & \multicolumn{3}{c}{$\tau_{MS}^{}$} \\
\hline \hline
 1 & scLH23t-mBR & $ 2.25 $ & $ -0.31 $ & $ 3.40 $ & 1 & $ 2.39 $ & $ 0.36 $ & $ 3.78 $ \\
 2 & scLH22t & $ 2.35 $ & $ -0.29 $ & $ 3.27 $ & 2 & $ 2.47 $ & $ 0.47 $ & $ 3.62 $ \\
 3 & LH20t & $ 2.48 $ & $ 0.46 $ & $ 3.71 $ & 4 & $ 2.71 $ & $ 1.15 $ & $ 4.09 $ \\
 4 & scLH23t-mBR-P & $ 2.52 $ & $ -0.41 $ & $ 3.81 $ & 3 & $ 2.69 $ & $ 0.27 $ & $ 4.20 $ \\
 5 & scLH22ta & $ 2.67 $ & $ 0.24 $ & $ 3.64 $ & 5 & $ 2.83 $ & $ 0.66 $ & $ 3.78 $ \\
 6 & LH14t-calPBE & $ 3.02 $ & $ 1.28 $ & $ 4.06 $ & 8 & $ 3.21 $ & $ 2.06 $ & $ 4.21 $ \\
 7 & $\omega$LH22t & $ 3.09 $ & $ 0.12 $ & $ 4.06 $ & 10 & $ 3.49 $ & $ 1.16 $ & $ 4.73 $ \\
 8 & BHandHLYP & $ 3.13 $ & $ 2.17 $ & $ 4.61 $ & 7 & $ 3.13 $ & $ 2.17 $ & $ 4.61 $ \\
 9 & $\omega$B97X-V & $ 3.23 $ & $ 2.53 $ & $ 4.31 $ & 9 & $ 3.23 $ & $ 2.53 $ & $ 4.31 $ \\
 10 & LH20t nonCal & $ 3.31 $ & $ 0.23 $ & $ 4.39 $ & 14 & $ 3.68 $ & $ 1.74 $ & $ 4.97 $ \\
 11 & LH07t-SVWN & $ 3.55 $ & $ 0.18 $ & $ 4.46 $ & 13 & $ 3.68 $ & $ 2.01 $ & $ 4.72 $ \\
 12 & scLH21ct-SVWN-m & $ 3.61 $ & $ -2.73 $ & $ 3.46 $ & 6 & $ 3.07 $ & $ -0.15 $ & $ 3.99 $ \\
 13 & LH12ct-SsirPW92 & $ 3.74 $ & $ -1.89 $ & $ 4.39 $ & 11 & $ 3.59 $ & $ 0.14 $ & $ 4.63 $ \\
 14 & $\omega$B97M-V & $ 3.87 $ & $ 1.40 $ & $ 5.03 $ & 12 & $ 3.62 $ & $ 0.43 $ & $ 4.70 $ \\
 15 & LH12ct-SsifPW92 & $ 4.25 $ & $ -2.70 $ & $ 4.67 $ & 15 & $ 3.93 $ & $ -0.67 $ & $ 4.94 $ \\
 16 & B97M-V & $ 4.78 $ & $ 3.50 $ & $ 5.79 $ & 16 & $ 5.19 $ & $ 4.13 $ & $ 5.48 $ \\
 17 & B3LYP5 & $ 5.44 $ & $ 4.55 $ & $ 5.93 $ & 17 & $ 5.44 $ & $ 4.55 $ & $ 5.93 $ \\
 18 & LH07s-SVWN & $ 5.84 $ & $ 3.12 $ & $ 7.24 $ & 18 & $ 5.84 $ & $ 3.12 $ & $ 7.24 $ \\
 19 & MN15 & $ 6.00 $ & $ 5.02 $ & $ 6.60 $ & 29 & $ 11.47 $ & $ 10.46 $ & $ 12.63 $ \\
 20 & TPSSh & $ 6.82 $ & $ 6.73 $ & $ 6.22 $ & 23 & $ 7.22 $ & $ 7.09 $ & $ 5.97 $ \\
 21 & mPSTS-noa2 & $ 6.85 $ & $ 6.83 $ & $ 6.27 $ & 22 & $ 7.15 $ & $ 7.13 $ & $ 5.99 $ \\
 22 & VSXC & $ 6.96 $ & $ 5.24 $ & $ 7.75 $ & 20 & $ 7.07 $ & $ 5.50 $ & $ 7.52 $ \\
 23 & LHJ14 & $ 6.98 $ & $ 5.62 $ & $ 6.96 $ & 21 & $ 7.13 $ & $ 5.56 $ & $ 7.55 $ \\
 24 & MN15-L & $ 7.07 $ & $ -6.84 $ & $ 6.04 $ & 19 & $ 6.55 $ & $ -5.25 $ & $ 6.81 $ \\
 25 & mPSTS-a1 & $ 7.10 $ & $ 7.04 $ & $ 6.34 $ & 24 & $ 7.42 $ & $ 7.37 $ & $ 6.07 $ \\
 26 & PW6B95 & $ 7.60 $ & $ 7.26 $ & $ 7.20 $ & 26 & $ 8.32 $ & $ 8.00 $ & $ 8.10 $ \\
 27 & M06-2X & $ 7.80 $ & $ 7.26 $ & $ 9.27 $ & 28 & $ 10.18 $ & $ 9.03 $ & $ 12.93 $ \\
 28 & TPSS & $ 7.83 $ & $ 7.46 $ & $ 6.94 $ & 25 & $ 8.24 $ & $ 7.88 $ & $ 6.77 $ \\
 29 & $\tau$-HCTH & $ 10.20 $ & $ 9.68 $ & $ 8.06 $ & 27 & $ 9.74 $ & $ 9.22 $ & $ 7.83 $ \\
 30 & M06-L & $ 12.93 $ & $ 12.86 $ & $ 11.32 $ & 30 & $ 12.53 $ & $ 12.49 $ & $ 9.30 $ \\
 31 & M06 & $ 15.06 $ & $ 14.78 $ & $ 13.12 $ & 31 & $ 13.37 $ & $ 13.14 $ & $ 12.98 $ \\
\hline
\end{tabular}
\caption{Mean absolute errors (MAEs), mean errors (MEs), and standard deviations (STDs) of the errors in the magnetizabilities of the 27 studied molecules in units of $ 10^{-30} $ J/T$^2$ from the CCSD(T) reference values with the studied functionals}
\label{tab:maemestd}
\end{table*}

\begin{figure*}
\begin{center}
\subfigure[LH functionals]{
\includegraphics[width=0.33\linewidth]{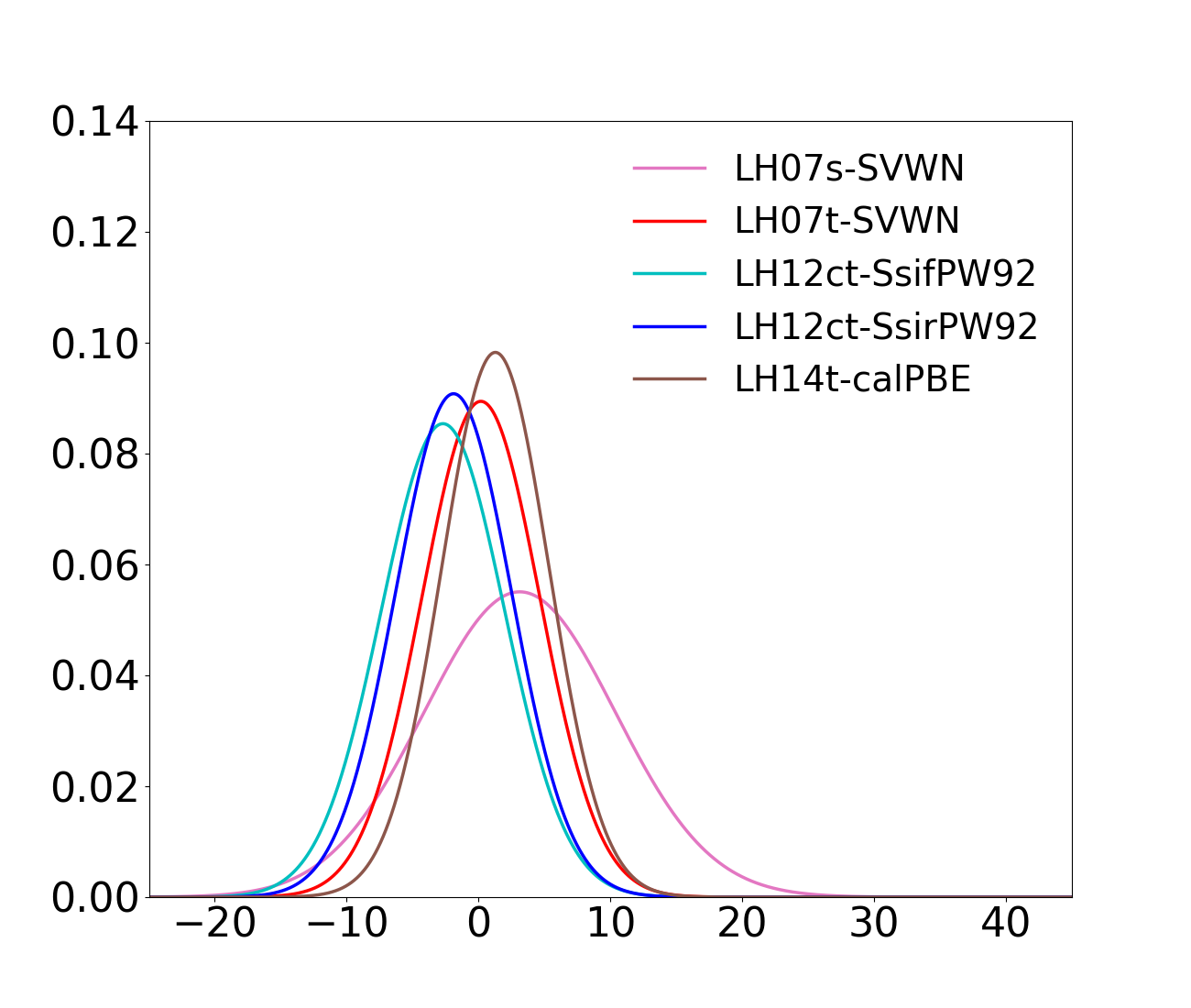}
\includegraphics[width=0.33\linewidth]{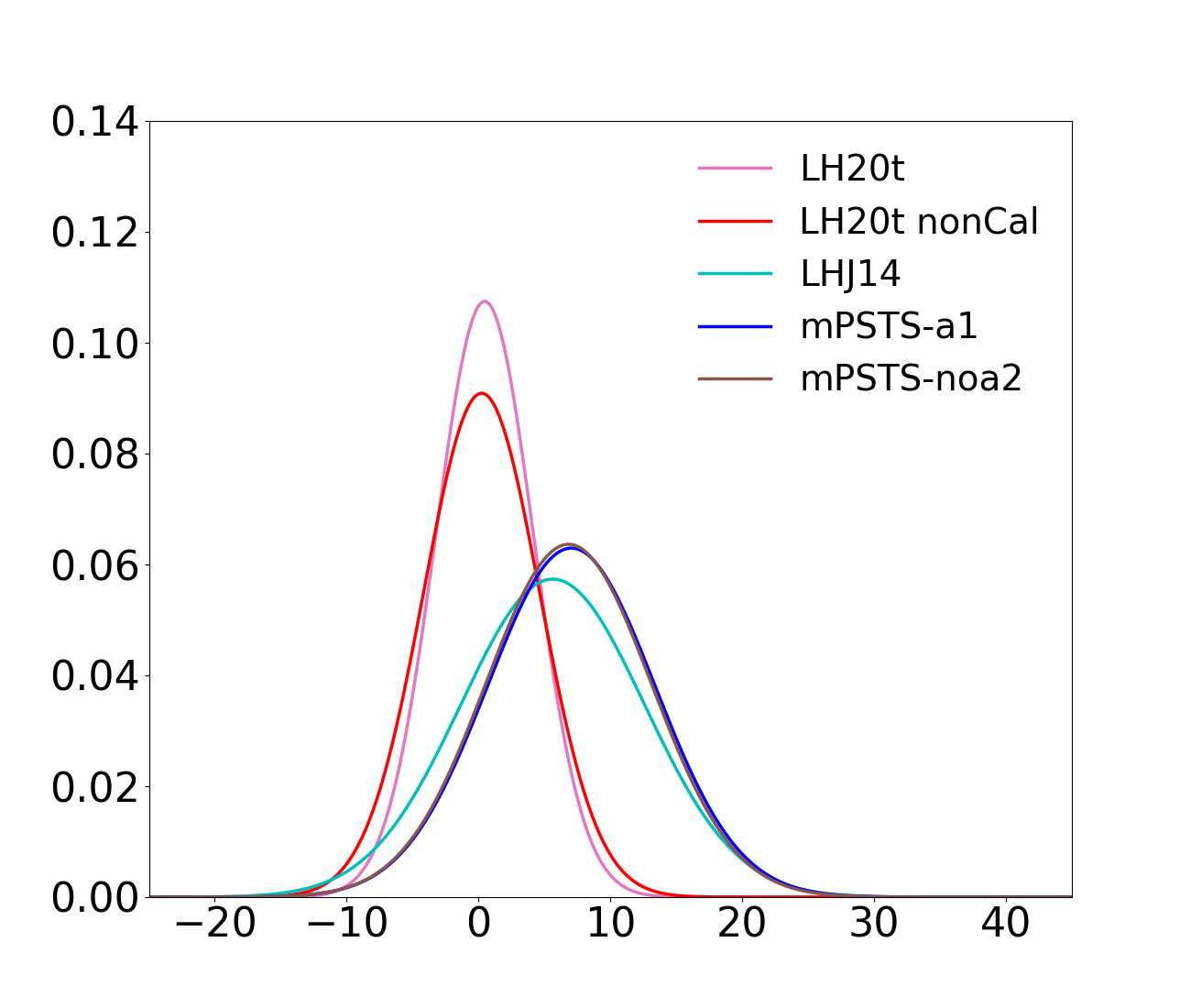}
}
\subfigure[mGGA functionals]{
\includegraphics[width=0.33\linewidth]{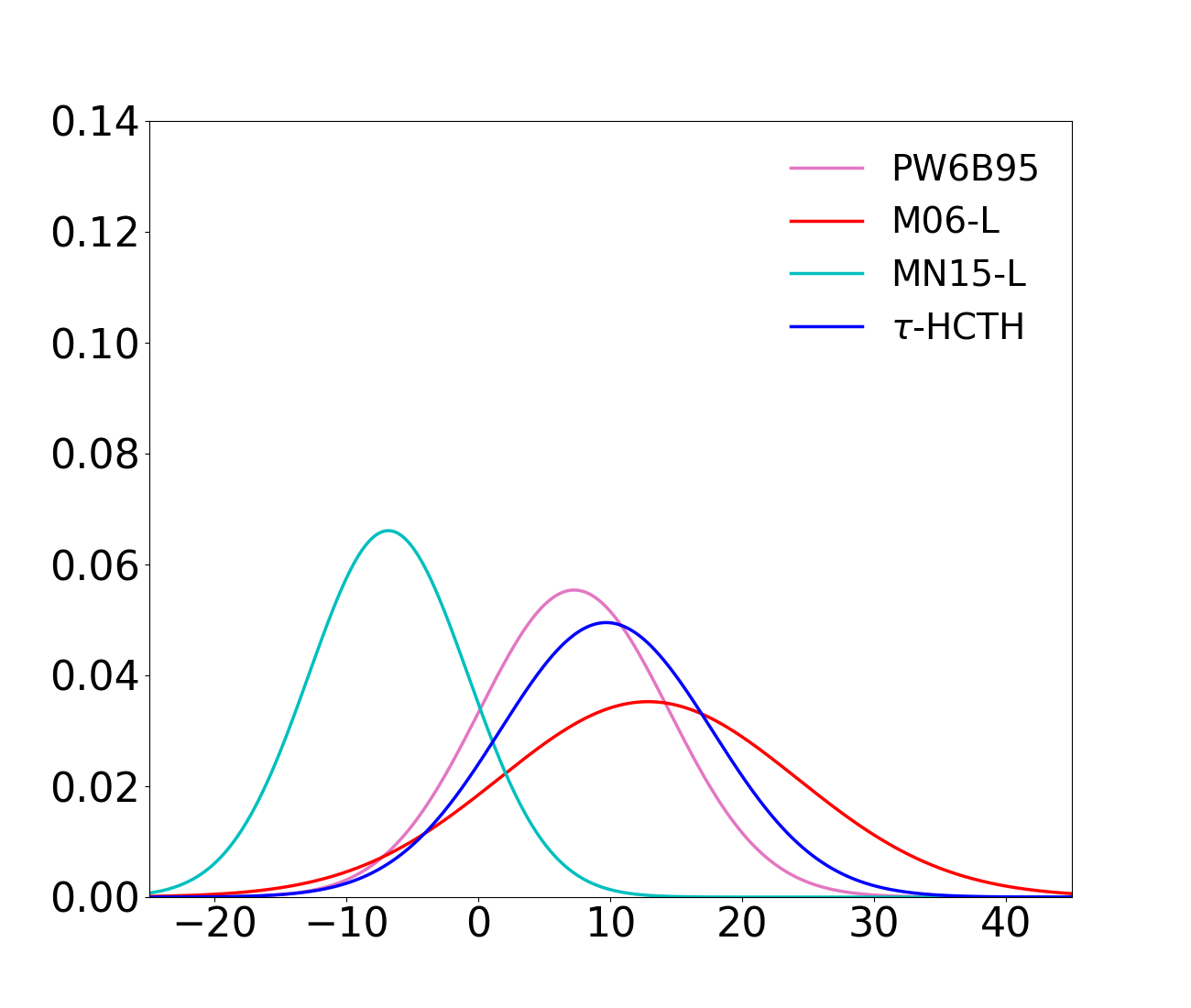}
\includegraphics[width=0.33\linewidth]{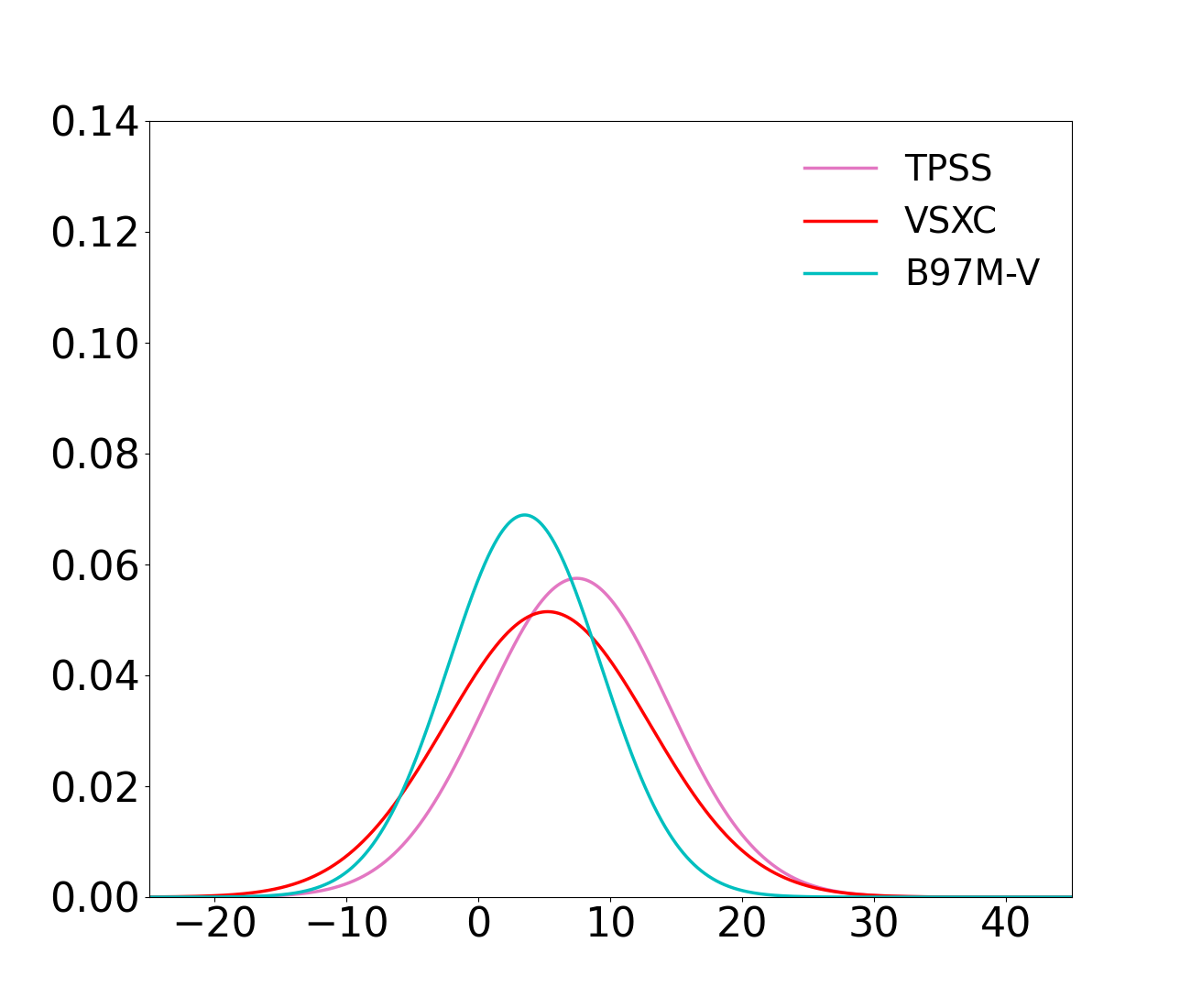}
}
\subfigure[GH functionals]{
\includegraphics[width=0.33\linewidth]{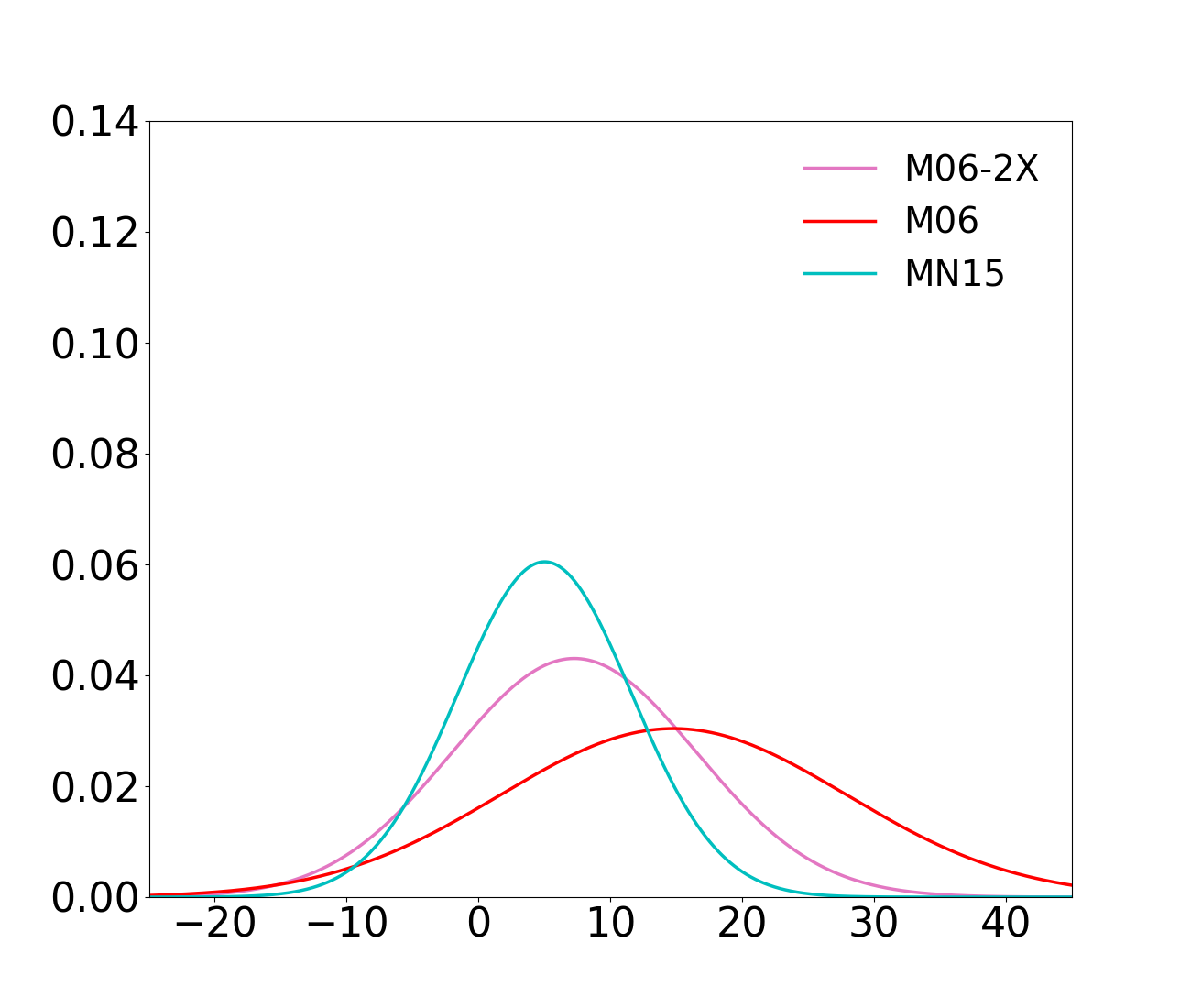}
\includegraphics[width=0.33\linewidth]{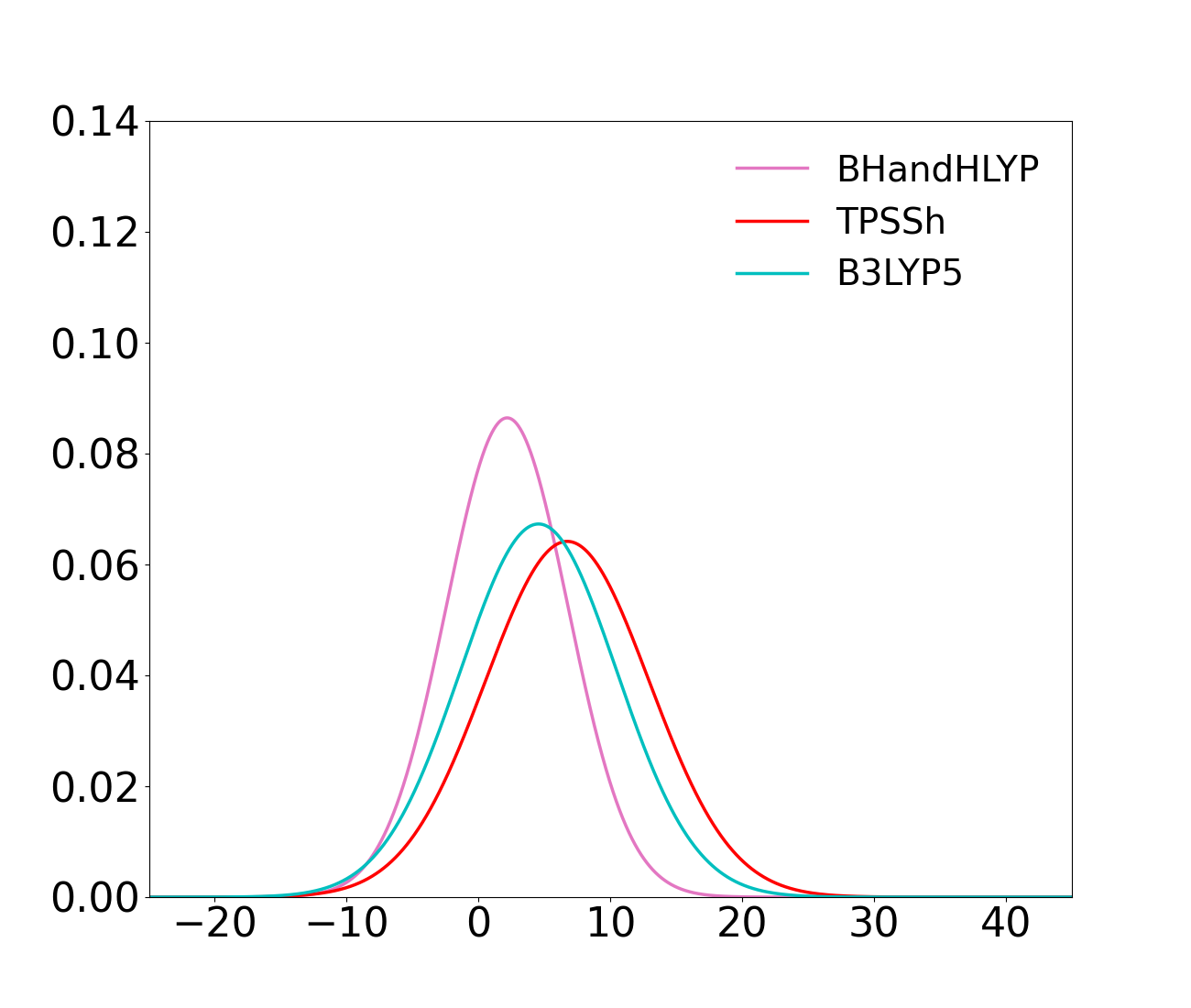}
}
\subfigure[RSH and scLH functionals]{
\includegraphics[width=0.33\linewidth]{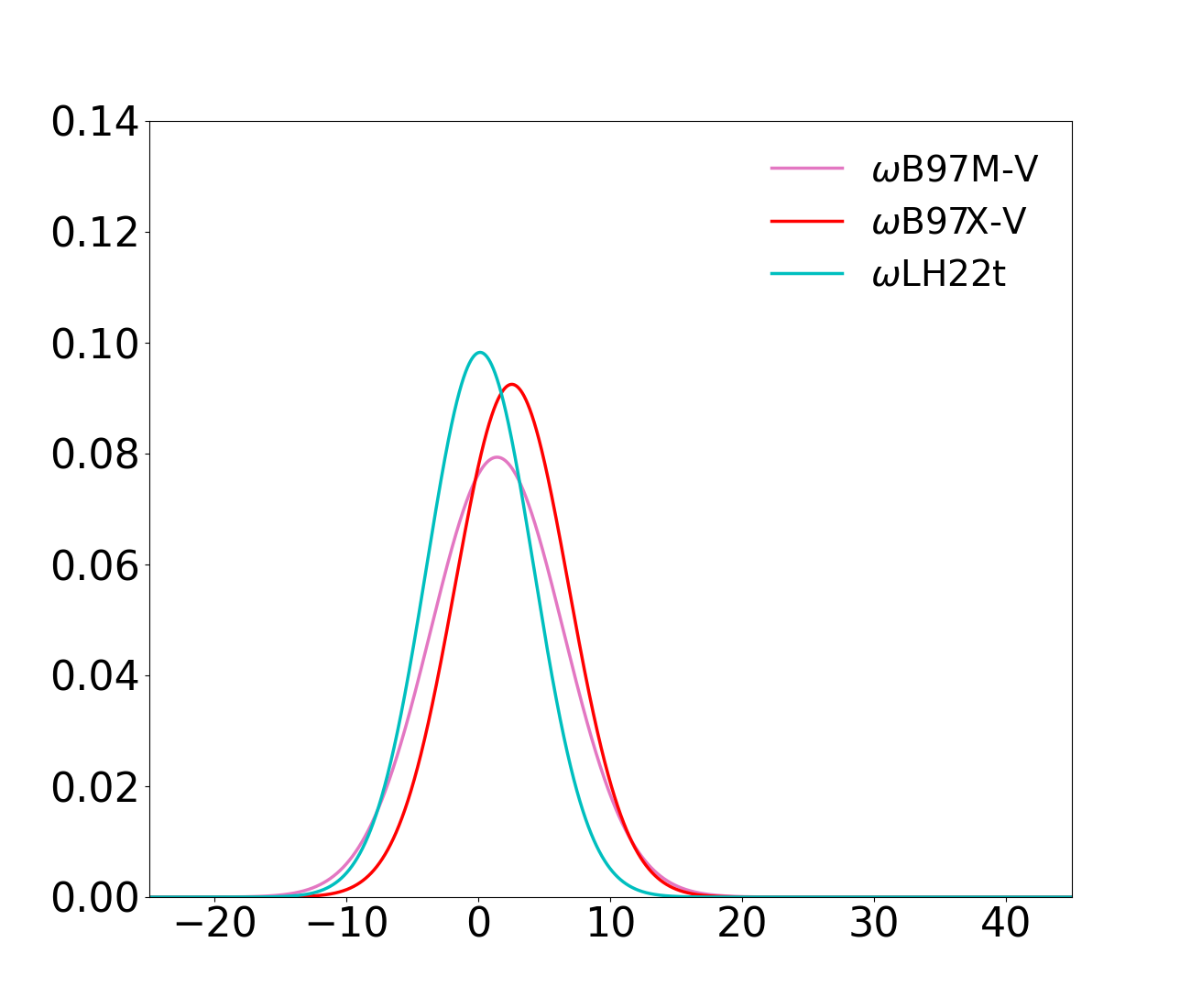}
\includegraphics[width=0.33\linewidth]{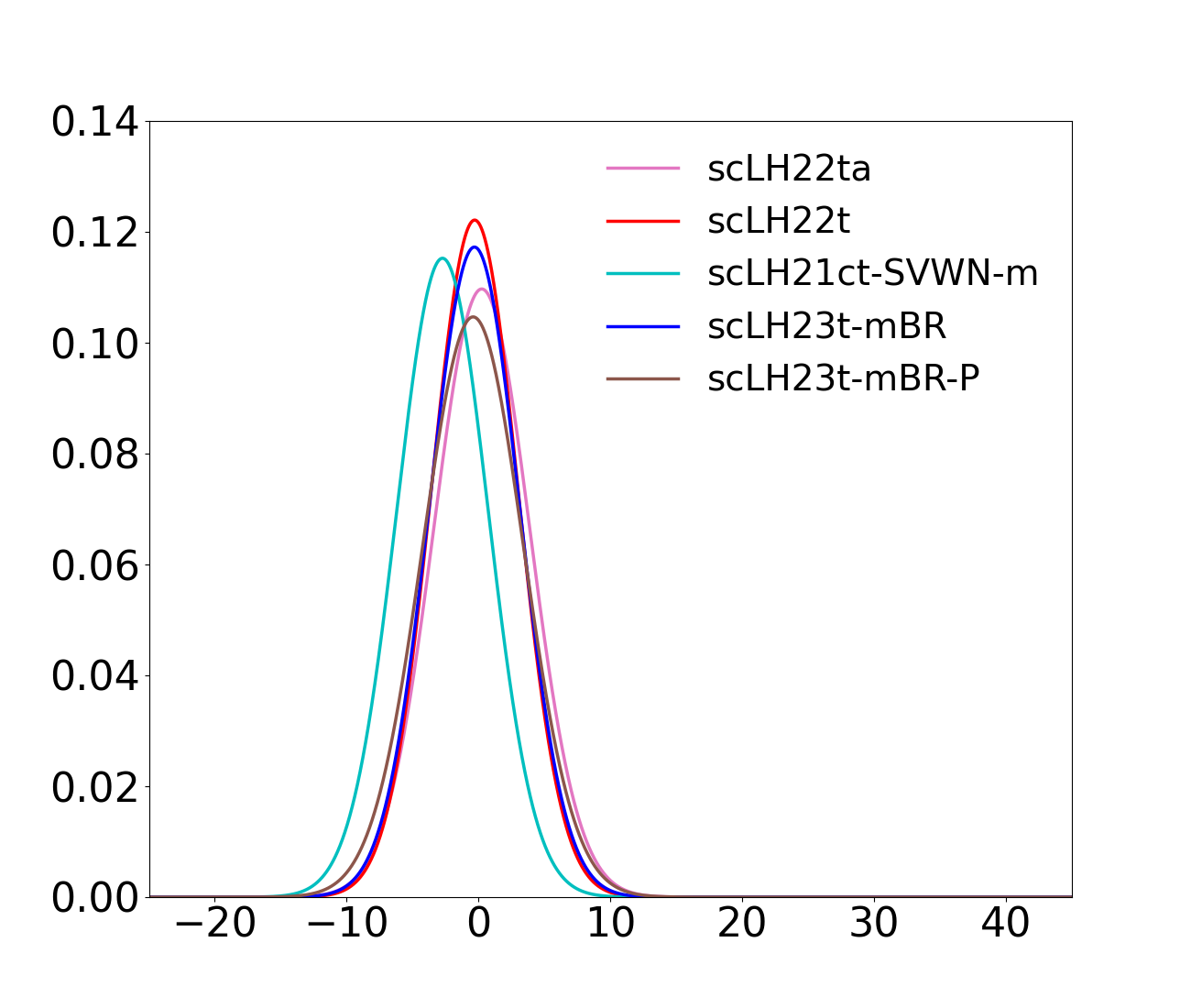}
}
\caption{Normal distributions of the magnetizability data calculated with
the Dobson formulation of $\tau$.
}
\label{fig:figure1}
\end{center}
\end{figure*}

In the mean absolute error (MAE) analysis, the largest effects of using
$\tau_\text{D}$ instead of $\tau_\text{MS}$ are obtained for some Minnesota
functionals, \textit{e.g.}, \jouletesla{-5.47} for MN15, \jouletesla{-2.38} for
M06-2X, and \jouletesla{+1.69} for M06.  Other functionals that exhibit large
(more than \jouletesla{0.5}) effects on the MAEs include PW6B95
(\jouletesla{-0.72}) and MN15-L (\jouletesla{+0.52}). We note here that many
Minnesota functionals, including M06 and M06-2X, have been recently found to be
numerically ill-behaved, while MN15 and MN15-L appear to behave
better.\cite{Lehtola2023_JCTC_2502, Lehtola2023_JPCA_4180}

The MAE values suggest that ensuring proper gauge invariance may either improve
or worsen the agreement with the CCSD(T) reference data. Some of us have found
a similar behavior for NMR chemical shifts,\cite{Schattenberg2021_JCTC_7602,
Schattenberg2021_JCTC_273} and attributed it to massive error compensation in
some of the cases.

However, in most cases the differences between the MS and Dobson
formulations are small, and the rankings of the best-performing
functionals are not affected very much by switching from using the MS
expression to the Dobson expression. Some exceptions do occur; for
instance the change of \jouletesla{-0.40} going from $\tau_\text{MS}$
to $\tau_\text{D}$ leads to an improved ranking by three positions for
$\omega$LH22t.

Interesting differences can be seen between the behavior of some
first-generation LH functionals based on LSDA exchange-energy densities
(\textit{e.g.}, LH12ct-SsirPW92, LH12ct-SsifPW92, and the related
scLH21ct-SVWN-m) and that of the more advanced functionals (LH, scLH, and the $\omega$LH22t RSLH functional). The accuracy of functionals of the
former type deteriorates somewhat after switching from the computationally
convenient $\tau_\text{MS}$ to the physically more correct $\tau_\text{D}$,
while the accuracy of the latter type of functionals improves when
$\tau_\text{D}$ is used.

A closer analysis of the origin of the differences between the
MS and Dobson formulations of $\tau$ is beyond the scope of
this work. However, a recent study of the Dobson-based gauge-invariance
contributions to TDDFT excitation energies has been able to link the magnitude
and even the sign of the effect to the way in which $\tau$ enters the
enhancement factor of the mGGA functionals
and other $\tau$-dependent functionals.\cite{Grotjahn2022_JCP_111102}

LH20t and its sc-corrected extensions scLH22t and scLH23t-mBR occupy ranks 3, 2
and 1, respectively, with the scLHs improving slightly over their parent LH
functional.
The reduced MAE of scLH23t-mBR and scLH22t as compared to LH20t arises from
some of the molecules that are expected to exhibit larger static correlation
effects, as evidenced by larger deviations between CCSD(T) and CCSD results in
Ref.  \citenum{Lutnes2009_JCP_144104}. Apart from the true static correlation
case \ce{O3} (see below), scLH22t gives notable improvements of more than
\jouletesla{2} for \ce{PN} and \ce{SO2}. scLH23t-mBR gives large improvements
for \ce{PN} and \ce{HCP}. The performance of both functionals would be even
more impressive if not for the somewhat larger deviations of \jouletesla{-6.4}
(scLH22t) and \jouletesla{-5.3} for the \ce{LiH} molecule as compared to the
LH20t value of \jouletesla{+3.2}. scLH23t-mBR-P, with its Pad\'e-based
$q_{AC}^{ }$  also improves particularly on \ce{PN} and \ce{HCP} but
deteriorates on \ce{LiH} (\jouletesla{-10.6}), hampering the functionals's
overall statistical performance compared to the other two scLHs and leaves it
slightly behind LH20t in the overall ranking.  Effects of the sc-corrections
are much smaller for the other molecules, which is consistent with an efficient
damping of the sc-factor for weakly correlated systems. It is presently unclear
why the effects of the sc-corrections are below \jouletesla{1} for several
other systems with larger differences between CCSD and CCSD(T) (\ce{H2CO},
\ce{OF2}, \ce{HOF}).

The two best-performing functionals in \citerefs{Lehtola2021_JCTC_1457} and
\citenum{Lehtola2021_JCTC_4629} (BHandHLYP and $\omega$B97X-V) are ranked
8$^\text{th}$ and 9$^\text{th}$ when considering the Dobson formulation of
$\tau$.  Strikingly, several LH and scLH functionals, as well as the
$\omega$LH22t RSLH functional occupy the seven top positions in the ranking,
and several further LH functionals follow in the top 15 of the ranking
(\cref{tab:maemestd}).  A similar trend was observed earlier when considering a
smaller selection of LH functionals.\cite{Holzer2021_JCTC_2928} However,
somewhat different conclusions were then reached because the experimental data
used as reference values in that work exhibit very large error bars and are
therefore ill-suited for benchmarking purposes.

Individual changes from LH20t to scLH22ta which lacks the damping factor are
overall somewhat more pronounced and lead to a slightly larger MAE for the scLH
functionals, indicating some deterioration of the accuracy of the
magnetizability for more weakly correlated systems, which places scLH22ta
behind LH20t, but scLH22ta is still ranked $5^\mathrm{th}$ best.

Further top-performing functionals include LH14t-calPBE (rank 6) and
the $\omega$LH22t RSLH (rank 7), but their MAE in the 3.0-3.3 $\times
10^{-30}$ J/T$^{2}$ range is already comparable to those of the
best-performing functionals from \citerefs{Lehtola2021_JCTC_1457} and
\citenum{Lehtola2021_JCTC_4629} (BHandHLYP, \mbox{$\omega$B97X-V}, and
\mbox{CAM-QTP} functionals).  In agreement with
\citerefs{Lehtola2021_JCTC_1457} and \citenum{Lehtola2021_JCTC_4629},
\mbox{B97M-V} remains the highest-ranked non-hybrid functional in the
evaluations, its MAE is reduced from \jouletesla{5.19} to
\jouletesla{4.78} by using the Dobson formulation of $\tau$. The
normal distributions of the magnetizabilities of the studied
functionals with the Dobson formula for $\tau$ are shown in
\cref{fig:figure1}; analogous plots for the MS formula for $\tau$ are
given in the SI, along with violin plots of the errors for both
descriptions, both with and without the inclusion of \ce{O3} in the
data set.

\subsection{Accuracy on \ce{O3}}
\label{sec:O3}

As was already discussed above, due to its large static correlation effects,
\ce{O3} has been excluded from statistical evaluations in previous density
functional assessments.\cite{Lutnes2009_JCP_144104, Lehtola2021_JCTC_1457,
Lehtola2021_JCTC_4629} Its large deviations would otherwise dominate the
statistics and it is unclear whether the CCSD(T) reference value is
sufficiently accurate for this molecule, given that the inclusion of the
perturbational triples (T) contributions reduces the magnetizability by more
than \jouletesla{45} as compared to the CCSD value.\cite{Lutnes2009_JCP_144104}
In the absence of experimental data, we may estimate a reasonable
range of values by also considering earlier complete active space
self-consistent field (CASSCF) GIAO
(\jouletesla{97.8})\cite{Ruud1997_JCP_10599} and multiconfigurational
(MC) individual gauge for localized orbitals (IGLO)
(\jouletesla{89.7})\cite{vanWuellenThesis} results. These values are
somewhat smaller than the CCSD(T) reference value of
\jouletesla{121.5} and suggest that magnetizabilities around
\jouletesla{100} with an error margin of about \jouletesla{\pm 20}
seem to define the most likely range. We review the various DFT
results in light of this range in \cref{tab:ozone}.

Most functionals overestimate the magnetizability of \ce{O3}
dramatically. Among the standard functionals studied here and in
\citerefs{Lehtola2021_JCTC_1457} and \citenum{Lehtola2021_JCTC_4629},
the specialized GGA functionals like KT1, KT2, and KT3 get closest to
the reference value, with KT1 providing the smallest value of
\jouletesla{131.9}. These DFAs are not good performers in general
according to the previous magnetizability
benchmark,\cite{Lehtola2021_JCTC_1457, Lehtola2021_JCTC_4629} and
would enter at ranks 19, 21, and 31 with MAE of \jouletesla{5.87},
\jouletesla{6.42} and \jouletesla{9.19}, respectively, in the
statistical evaluation of \cref{tab:maemestd}.
Other standard GGA functionals like BLYP or BP86 give
magnetizabilities of \ce{O3} around
\jouletesla{180}.\cite{Lehtola2021_JCTC_1457} Similar results for
simple GGA functionals were also obtained in the original evaluation
of \citeauthor{Lutnes2009_JCP_144104}.\cite{Lutnes2009_JCP_144104} However, these functionals perform
poorly for the entire test set, which is why we will not discuss them
further here. Some mGGA functionals like M06-L and B97M-V also attain
closer agreement when the $\tau_\text{MS}$ prescription is used, but
using the more appropriate $\tau_\text{D}$ results in larger
magnetizability and leads to a poorer agreement. When the Dobson
formulation is used, most mGGA functionals perform comparably to the
simpler GGA functionals, with TPSS and $\tau$-HCTH giving the lowest
values of \jouletesla{173.5} and \jouletesla{178.5}, respectively.

\begin{table}
\footnotesize
\caption{Calculated isotropic magnetizabilities [$10^{-30}$ J/T$^{2}$] for ozone with various methods.}
\label{tab:ozone}
\begin{tabularx}{0.47\textwidth}{@{}L|l|rr@{}}
\hline
Functional& Type & \multicolumn{2}{c}{$\overline{\xi} [{\text{\ce{O3}}}]$}\\
				&	&	$\tau_\text{D}$		&	$\tau_\text{MS}$  \\
\hline\hline
B3LYP & GH, GGA			&   \multicolumn{2}{c}{238.5}  \\
BHandHLYP & GH, GGA 		&   \multicolumn{2}{c}{336.6}  \\
B97M-V	&mGGA			&189.8		&99.3   \\
M06-L	&mGGA			&244.2		&156.2  \\
MN15-L	&mGGA			&193.7		&63.6  	\\
PW6B95 &mGGA 			&261.4		&288.3  \\
$\tau$-HCTH &mGGA		&178.3		&177.0  \\
TPSS	&mGGA			&173.5		&151.1  \\
VSXC     &mGGA				&193.9		&155.3  \\
M06       &GH, mGGA		&417.8		&413.4  \\
MN15   & GH, mGGA		&259.5		&570.8  \\
M06-2X &GH, mGGA		&328.1		&492.9  \\
TPSSh & GH, mGGA		&200.1		&176.9  \\
$\omega$B97X-V & RSH, GGA	&	\multicolumn{2}{c}  {251.2}  \\
$\omega$B97M-V & RSH, mGGA	&239.8		&225.3  \\
LH07s-SVWN & LH			&	\multicolumn{2}{c} {256.0} \\
LH07t-SVWN & LH 		&211.2		&207.3  \\
\mbox{LH12ct-SsifPW92} & LH		&221.6		&227.4  \\
\mbox{LH12ct-SsirPW92} & LH		&218.2		&220.2  \\
LH14t-calPBE   & LH 		&207.5		&207.2  \\
LH20t	& LH			&211.4		&223.1  \\
LH20t nonCal & LH	  	&217.3		&229.9  \\
LHJ14	   & LH 		&213.6		&244.3  \\
mPSTS-a1   & LH			&198.9		&175.5  \\
mPSTS-noa2 & LH			&207.5		&183.8  \\
\mbox{scLH22ct-SVWN-m} & scLH	&112.9		&90.8  	\\
scLH22ta & scLH			&101.0		&90.8  	\\
scLH22t	& scLH			&112.2		&109.5  \\
scLH23t-mBR & scLH 		&120.4		&	121.4		\\
scLH23t-mBR-P & scLH 		&134.4		& 136.5			\\
$\omega$LH22t  & RSLH		&224.9		&240.9  \\
\hline
\mbox{CCSD(T)-GIAO$^a$}& WFT	& \multicolumn{2}{c} {121.5}   \\
CASSCF-GIAO$^b$&  WFT	&	\multicolumn{2}{c} {97.8}  \\
MC-IGLO$^c$ &  WFT	&	\multicolumn{2}{c}	{89.7}  \\
\hline\hline
\multicolumn{4}{l}{\mbox{\scriptsize{$^a$ \citeref{Lehtola2021_JCTC_1457}; $^b$ \citeref{Ruud1997_JCP_10599}; $^c$ \citeref{vanWuellenThesis}}}}
\end{tabularx}
\end{table}

Including exact exchange in GH and RSH functionals significantly increases the
calculated magnetizability of ozone, and thereby deteriorates the agreement
with the reference value, in some cases they are  more than \jouletesla{300}
(BHandHLYP, M06, M06-2X).  We find a similar trend for all LH functionals
without sc-corrections as well as for the $\omega$LH22t RSLH functional that
yield values larger than \jouletesla{200}.  The strikingly good performance of
various scLH functionals is thus particularly notable: scLH22ta without damping
factor in the sc-corrections gives the lowest value, results for the other
sc-functionals are in the range of \jouletesla{112-135}.

Given that four of the five scLHs are also among the five
best-performing functionals for the entire test set, these results are
significant in suggesting that the scLH functionals to some extent
indeed escape the usual zero-sum game between delocalization errors
and strong-correlation errors, where larger EXX admixtures improve on
the former but deteriorate the latter.\cite{Janesko2017_JPCL_4314}
Such an ``escape'' has been found recently for fractional spin errors
and the related spin-restricted dissociation curves of
diatomics.\cite{Wodynski2022_JCTC_6111, Wodynski2023_JCP_244117} It is
gratifying to see this here for a very different property.

\section{Conclusions}
\label{sec:conclusions}

This work extends in two directions previous studies of DFT
functionals for the computation of molecular magnetizabilities. We
considered using a gauge-independent local kinetic energy $\tau$
ingredient in a wide variety of mGGA functionals within Dobson's
current-DFT formalism ($\tau_\text{D}$), which had not been considered
in as much detail so far.  We examined the effects of the Dobson
formalism by comparing the obtained magnetizabilities to values
calculated with the Maximoff--Scuseria (MS) formulation
($\tau_\text{MS}$) used in previous works.  We also extended the
assessment to local hybrid functionals \textit{i.e.}, with
position-dependent exact-exchange admixtures, in particular to their
recent strong-correlation corrected and range-separated variants.

Regarding gauge invariant formulations of $\tau$, we find that going
from the previously used, computationally convenient $\tau_\text{MS}$
to the more physically correct $\tau_\text{D}$ leads in some cases to
dramatic changes in the magnetizability, while in other cases the
differences between $\tau_\text{D}$ and $\tau_\text{MS}$ are
small. The largest effects are seen for some of the highly
parameterized mGGA and mGGA hybrid functionals from the Minnesota
group, whose numerical behavior has also been recently investigated
and found wanting for many functionals.\cite{Lehtola2023_JPCA_4180,
  Lehtola2023_JCTC_2502}

While $\tau_\text{D}$ leads to improved agreement with the CCSD(T) reference
data for some functionals, it also deteriorates the agreement for other
functionals. Notably, the effects of making $\tau$ gauge-invariant by the
Dobson procedure tend to be smaller for the overall better-performing
functionals, which include many local and range-separated hybrid functionals.

The overall statistical evaluation of a wide variety of different functionals
provided evidence that local hybrid functionals can yield particularly
accurate magnetizabilities. Indeed, the seven best-performing functionals in
the present evaluation are newer local hybrid functionals, their
strong-correlation corrected variants, and the recently reported
range-separated local hybrid functional $\omega$LH22t.  The overall statistical
improvement compared to the so far best-performing functionals is moderate but
notable, \textit{e.g.}, MAEs of \jouletesla{2.25}, \jouletesla{2.35} and
\jouletesla{2.48} for scLH23t-mBR, scLH22t and LH20t, respectively, compared to
\jouletesla{3.11} for BHandHLYP and \jouletesla{3.23} for $\omega$B97X-V.

The most striking result is the dramatic improvement obtained with several of
the strong-correlation corrected local hybrid functionals for the
static-correlation case \ce{O3}.  Importantly, this improvement is achieved
while retaining the overall highest accuracy for the weakly correlated systems
relevant for the statistical evaluations. This observation for a totally
different property than evaluated so far for such functionals is a further
indication that strong-correlation corrected local hybrid functionals offer an
escape from the usual zero-sum game between achieving low fractional charge
errors and low fractional spin errors.

\begin{acknowledgement}

This work has been supported by the Academy of Finland through project
numbers 340583, 350282, and 353749, by Magnus Ehrnrooth Foundation,
the Finnish Society of Sciences and Letters and by the Swedish
Cultural Foundation in Finland. We acknowledge computational resources
from CSC---IT Center for Science, Finland. Work in Berlin has been
supported by DFG project KA1187/15-1.

\end{acknowledgement}

\begin{suppinfo}
Magnetizabilities, normal distributions and violin plots of the error
distributions of the magnetizabilities calculated with the
Maximoff--Scuseria and Dobson expressions of $\tau$. Equations
necessary to implement the magnetic field derivatives for scLHs and
RSLHs.

\end{suppinfo}
\clearpage

\bibliography{literature,libxc,susi,caspar}

\providecommand{\latin}[1]{#1}
\makeatletter
\providecommand{\doi}
  {\begingroup\let\do\@makeother\dospecials
  \catcode`\{=1 \catcode`\}=2 \doi@aux}
\providecommand{\doi@aux}[1]{\endgroup\texttt{#1}}
\makeatother
\providecommand*\mcitethebibliography{\thebibliography}
\csname @ifundefined\endcsname{endmcitethebibliography}
  {\let\endmcitethebibliography\endthebibliography}{}
\begin{mcitethebibliography}{142}
\providecommand*\natexlab[1]{#1}
\providecommand*\mciteSetBstSublistMode[1]{}
\providecommand*\mciteSetBstMaxWidthForm[2]{}
\providecommand*\mciteBstWouldAddEndPuncttrue
  {\def\EndOfBibitem{\unskip.}}
\providecommand*\mciteBstWouldAddEndPunctfalse
  {\let\EndOfBibitem\relax}
\providecommand*\mciteSetBstMidEndSepPunct[3]{}
\providecommand*\mciteSetBstSublistLabelBeginEnd[3]{}
\providecommand*\EndOfBibitem{}
\mciteSetBstSublistMode{f}
\mciteSetBstMaxWidthForm{subitem}{(\alph{mcitesubitemcount})}
\mciteSetBstSublistLabelBeginEnd
  {\mcitemaxwidthsubitemform\space}
  {\relax}
  {\relax}

\bibitem[Hohenberg and Kohn(1964)Hohenberg, and Kohn]{Hohenberg1964_PR_864}
Hohenberg,~P.; Kohn,~W. Inhomogeneous Electron Gas. \emph{Phys. Rev.}
  \textbf{1964}, \emph{136}, B864--B871\relax
\mciteBstWouldAddEndPuncttrue
\mciteSetBstMidEndSepPunct{\mcitedefaultmidpunct}
{\mcitedefaultendpunct}{\mcitedefaultseppunct}\relax
\EndOfBibitem
\bibitem[Kohn and Sham(1965)Kohn, and Sham]{Kohn1965_PR_1133}
Kohn,~W.; Sham,~L.~J. Self-Consistent Equations Including Exchange and
  Correlation Effects. \emph{Phys. Rev.} \textbf{1965}, \emph{140},
  A1133--A1138\relax
\mciteBstWouldAddEndPuncttrue
\mciteSetBstMidEndSepPunct{\mcitedefaultmidpunct}
{\mcitedefaultendpunct}{\mcitedefaultseppunct}\relax
\EndOfBibitem
\bibitem[Helgaker \latin{et~al.}(1999)Helgaker, Jaszu{\'{n}}ski, and
  Ruud]{Helgaker1999_CR_293}
Helgaker,~T.; Jaszu{\'{n}}ski,~M.; Ruud,~K. Ab Initio Methods for the
  Calculation of {NMR} Shielding and Indirect Spin--Spin Coupling Constants.
  \emph{Chem. Rev.} \textbf{1999}, \emph{99}, 293--352\relax
\mciteBstWouldAddEndPuncttrue
\mciteSetBstMidEndSepPunct{\mcitedefaultmidpunct}
{\mcitedefaultendpunct}{\mcitedefaultseppunct}\relax
\EndOfBibitem
\bibitem[Gauss and Stanton(2002)Gauss, and Stanton]{Gauss2002_ACP_355}
Gauss,~J.; Stanton,~J.~F. Electron-Correlated Approaches for the Calculation of
  {NMR} Chemical Shifts. \emph{Adv. Chem. Phys.} \textbf{2002}, \emph{123},
  355--422\relax
\mciteBstWouldAddEndPuncttrue
\mciteSetBstMidEndSepPunct{\mcitedefaultmidpunct}
{\mcitedefaultendpunct}{\mcitedefaultseppunct}\relax
\EndOfBibitem
\bibitem[Keal \latin{et~al.}(2004)Keal, Tozer, and Helgaker]{Keal2004_CPL_374}
Keal,~T.~W.; Tozer,~D.~J.; Helgaker,~T. {GIAO shielding constants and indirect
  spin--spin coupling constants: performance of density functional methods}.
  \emph{Chem. Phys. Lett.} \textbf{2004}, \emph{391}, 374--379\relax
\mciteBstWouldAddEndPuncttrue
\mciteSetBstMidEndSepPunct{\mcitedefaultmidpunct}
{\mcitedefaultendpunct}{\mcitedefaultseppunct}\relax
\EndOfBibitem
\bibitem[Gauss \latin{et~al.}(2007)Gauss, Ruud, and
  K{\'{a}}llay]{Gauss2007_JCP_74101}
Gauss,~J.; Ruud,~K.; K{\'{a}}llay,~M. {Gauge-origin independent calculation of
  magnetizabilities and rotational g tensors at the coupled-cluster level.}
  \emph{J. Chem. Phys.} \textbf{2007}, \emph{127}, 074101\relax
\mciteBstWouldAddEndPuncttrue
\mciteSetBstMidEndSepPunct{\mcitedefaultmidpunct}
{\mcitedefaultendpunct}{\mcitedefaultseppunct}\relax
\EndOfBibitem
\bibitem[Lutn{\ae}s \latin{et~al.}(2009)Lutn{\ae}s, Teale, Helgaker, Tozer,
  Ruud, and Gauss]{Lutnes2009_JCP_144104}
Lutn{\ae}s,~O.~B.; Teale,~A.~M.; Helgaker,~T.; Tozer,~D.~J.; Ruud,~K.;
  Gauss,~J. Benchmarking density-functional-theory calculations of rotational
  $g$ tensors and magnetizabilities using accurate coupled-cluster
  calculations. \emph{J. Chem. Phys.} \textbf{2009}, \emph{131}, 144104\relax
\mciteBstWouldAddEndPuncttrue
\mciteSetBstMidEndSepPunct{\mcitedefaultmidpunct}
{\mcitedefaultendpunct}{\mcitedefaultseppunct}\relax
\EndOfBibitem
\bibitem[Teale \latin{et~al.}(2013)Teale, Lutn{\ae}s, Helgaker, Tozer, and
  Gauss]{Teale2013_JCP_24111}
Teale,~A.~M.; Lutn{\ae}s,~O.~B.; Helgaker,~T.; Tozer,~D.~J.; Gauss,~J.
  Benchmarking density-functional theory calculations of {NMR} shielding
  constants and spin-rotation constants using accurate coupled-cluster
  calculations. \emph{J. Chem. Phys.} \textbf{2013}, \emph{138}, 024111\relax
\mciteBstWouldAddEndPuncttrue
\mciteSetBstMidEndSepPunct{\mcitedefaultmidpunct}
{\mcitedefaultendpunct}{\mcitedefaultseppunct}\relax
\EndOfBibitem
\bibitem[Loibl and Sch{\"{u}}tz(2014)Loibl, and
  Sch{\"{u}}tz]{Loibl2014_JCP_24108}
Loibl,~S.; Sch{\"{u}}tz,~M. Magnetizability and rotational g tensors for
  density fitted local second-order {M}{\o}ller--{P}lesset perturbation theory
  using gauge-including atomic orbitals. \emph{J. Chem. Phys.} \textbf{2014},
  \emph{141}, 024108\relax
\mciteBstWouldAddEndPuncttrue
\mciteSetBstMidEndSepPunct{\mcitedefaultmidpunct}
{\mcitedefaultendpunct}{\mcitedefaultseppunct}\relax
\EndOfBibitem
\bibitem[Reimann \latin{et~al.}(2017)Reimann, Borgoo, Tellgren, Teale, and
  Helgaker]{Reimann2017_JCTC_4089}
Reimann,~S.; Borgoo,~A.; Tellgren,~E.~I.; Teale,~A.~M.; Helgaker,~T.
  {Magnetic-Field Density-Functional Theory (BDFT): Lessons from the Adiabatic
  Connection}. \emph{J. Chem. Theory Comput.} \textbf{2017}, \emph{13},
  4089--4100\relax
\mciteBstWouldAddEndPuncttrue
\mciteSetBstMidEndSepPunct{\mcitedefaultmidpunct}
{\mcitedefaultendpunct}{\mcitedefaultseppunct}\relax
\EndOfBibitem
\bibitem[Lehtola \latin{et~al.}(2021)Lehtola, Dimitrova, Fliegl, and
  Sundholm]{Lehtola2021_JCTC_1457}
Lehtola,~S.; Dimitrova,~M.; Fliegl,~H.; Sundholm,~D. Benchmarking
  Magnetizabilities with Recent Density Functionals. \emph{J. Chem. Theory
  Comput.} \textbf{2021}, \emph{17}, 1457--1468\relax
\mciteBstWouldAddEndPuncttrue
\mciteSetBstMidEndSepPunct{\mcitedefaultmidpunct}
{\mcitedefaultendpunct}{\mcitedefaultseppunct}\relax
\EndOfBibitem
\bibitem[Lehtola \latin{et~al.}(2021)Lehtola, Dimitrova, Fliegl, and
  Sundholm]{Lehtola2021_JCTC_4629}
Lehtola,~S.; Dimitrova,~M.; Fliegl,~H.; Sundholm,~D. Correction to
  "{Benchmarking} Magnetizabilities with Recent Density Functionals". \emph{J.
  Chem. Theory Comput.} \textbf{2021}, \emph{17}, 4629--4631\relax
\mciteBstWouldAddEndPuncttrue
\mciteSetBstMidEndSepPunct{\mcitedefaultmidpunct}
{\mcitedefaultendpunct}{\mcitedefaultseppunct}\relax
\EndOfBibitem
\bibitem[Perdew and Schmidt(2001)Perdew, and Schmidt]{Perdew2001_ACP_1}
Perdew,~J.~P.; Schmidt,~K. {Jacob}'s ladder of density functional
  approximations for the exchange-correlation energy. \emph{AIP Conf. Proc.}
  \textbf{2001}, \emph{577}, 1--20\relax
\mciteBstWouldAddEndPuncttrue
\mciteSetBstMidEndSepPunct{\mcitedefaultmidpunct}
{\mcitedefaultendpunct}{\mcitedefaultseppunct}\relax
\EndOfBibitem
\bibitem[Becke(1993)]{Becke1993_JCP_1372}
Becke,~A.~D. A new mixing of {Hartree}--{Fock} and local density-functional
  theories. \emph{J. Chem. Phys.} \textbf{1993}, \emph{98}, 1372--1377\relax
\mciteBstWouldAddEndPuncttrue
\mciteSetBstMidEndSepPunct{\mcitedefaultmidpunct}
{\mcitedefaultendpunct}{\mcitedefaultseppunct}\relax
\EndOfBibitem
\bibitem[Balasubramani \latin{et~al.}(2020)Balasubramani, Chen, Coriani,
  Diedenhofen, Frank, Franzke, Furche, Grotjahn, Harding, Hättig, Hellweg,
  Helmich-Paris, Holzer, Huniar, Kaupp, Marefat~Khah, Karbalaei~Khani, Müller,
  Mack, Nguyen, Parker, Perlt, Rappoport, Reiter, Roy, Rückert, Schmitz,
  Sierka, Tapavicza, Tew, van Wüllen, Voora, Weigend, Wodyński, and
  Yu]{TBM_Balasubramani2020_JCP_184107}
Balasubramani,~S.~G.; Chen,~G.~P.; Coriani,~S.; Diedenhofen,~M.; Frank,~M.~S.;
  Franzke,~Y.~J.; Furche,~F.; Grotjahn,~R.; Harding,~M.~E.; Hättig,~C.
  \latin{et~al.}  TURBOMOLE: Modular program suite for ab initio
  quantum-chemical and condensed-matter simulations. \emph{J. Chem. Phys.}
  \textbf{2020}, \emph{152}, 184107\relax
\mciteBstWouldAddEndPuncttrue
\mciteSetBstMidEndSepPunct{\mcitedefaultmidpunct}
{\mcitedefaultendpunct}{\mcitedefaultseppunct}\relax
\EndOfBibitem
\bibitem[TBM()]{TBM}
{TURBOMOLE V7.6 2021}, a development of {University of Karlsruhe} and
  {Forschungszentrum Karlsruhe GmbH}, 1989-2007, {TURBOMOLE GmbH}, since 2007;
  available from \\ {\tt https://www.turbomole.org}.\relax
\mciteBstWouldAddEndPunctfalse
\mciteSetBstMidEndSepPunct{\mcitedefaultmidpunct}
{}{\mcitedefaultseppunct}\relax
\EndOfBibitem
\bibitem[Maximoff and Scuseria(2004)Maximoff, and
  Scuseria]{Maximoff2004_CPL_408}
Maximoff,~S.~N.; Scuseria,~G.~E. Nuclear magnetic resonance shielding tensors
  calculated with kinetic energy density-dependent exchange-correlation
  functionals. \emph{Chem. Phys. Lett.} \textbf{2004}, \emph{390},
  408--412\relax
\mciteBstWouldAddEndPuncttrue
\mciteSetBstMidEndSepPunct{\mcitedefaultmidpunct}
{\mcitedefaultendpunct}{\mcitedefaultseppunct}\relax
\EndOfBibitem
\bibitem[Frisch \latin{et~al.}(2016)Frisch, Trucks, Schlegel, Scuseria, Robb,
  Cheeseman, Scalmani, Barone, Petersson, Nakatsuji, Li, Caricato, Marenich,
  Bloino, Janesko, Gomperts, Mennucci, Hratchian, Ortiz, Izmaylov, Sonnenberg,
  Williams-Young, Ding, Lipparini, Egidi, Goings, Peng, Petrone, Henderson,
  Ranasinghe, Zakrzewski, Gao, Rega, Zheng, Liang, Hada, Ehara, Toyota, Fukuda,
  Hasegawa, Ishida, Nakajima, Honda, Kitao, Nakai, Vreven, Throssell,
  Montgomery, Peralta, Ogliaro, Bearpark, Heyd, Brothers, Kudin, Staroverov,
  Keith, Kobayashi, Normand, Raghavachari, Rendell, Burant, Iyengar, Tomasi,
  Cossi, Millam, Klene, Adamo, Cammi, Ochterski, Martin, Morokuma, Farkas,
  Foresman, and Fox]{gaussian16}
Frisch,~M.~J.; Trucks,~G.~W.; Schlegel,~H.~B.; Scuseria,~G.~E.; Robb,~M.~A.;
  Cheeseman,~J.~R.; Scalmani,~G.; Barone,~V.; Petersson,~G.~A.; Nakatsuji,~H.
  \latin{et~al.}  Gaussian 16 {R}evision {C}.01. 2016; Gaussian Inc.
  Wallingford CT\relax
\mciteBstWouldAddEndPuncttrue
\mciteSetBstMidEndSepPunct{\mcitedefaultmidpunct}
{\mcitedefaultendpunct}{\mcitedefaultseppunct}\relax
\EndOfBibitem
\bibitem[Dobson(1993)]{Dobson1993_JCP_8870}
Dobson,~J.~F. {Alternative expressions for the Fermi hole curvature}. \emph{J.
  Chem. Phys.} \textbf{1993}, \emph{98}, 8870--8872\relax
\mciteBstWouldAddEndPuncttrue
\mciteSetBstMidEndSepPunct{\mcitedefaultmidpunct}
{\mcitedefaultendpunct}{\mcitedefaultseppunct}\relax
\EndOfBibitem
\bibitem[Schattenberg and Kaupp(2021)Schattenberg, and
  Kaupp]{Schattenberg2021_JCTC_1469}
Schattenberg,~C.~J.; Kaupp,~M. Effect of the Current Dependence of
  Tau-Dependent Exchange-Correlation Functionals on Nuclear Shielding
  Calculations. \emph{J. Chem. Theory Comput.} \textbf{2021}, \emph{17},
  1469--1479\relax
\mciteBstWouldAddEndPuncttrue
\mciteSetBstMidEndSepPunct{\mcitedefaultmidpunct}
{\mcitedefaultendpunct}{\mcitedefaultseppunct}\relax
\EndOfBibitem
\bibitem[Schattenberg and Kaupp(2021)Schattenberg, and
  Kaupp]{Schattenberg2021_JPCA_2697}
Schattenberg,~C.~J.; Kaupp,~M. Implementation and Validation of Local Hybrid
  Functionals with Calibrated Exchange-Energy Densities for Nuclear Shielding
  Constants. \emph{J. Phys. Chem. A} \textbf{2021}, \emph{125},
  2697--2707\relax
\mciteBstWouldAddEndPuncttrue
\mciteSetBstMidEndSepPunct{\mcitedefaultmidpunct}
{\mcitedefaultendpunct}{\mcitedefaultseppunct}\relax
\EndOfBibitem
\bibitem[Schattenberg and Kaupp(2021)Schattenberg, and
  Kaupp]{Schattenberg2021_JCTC_7602}
Schattenberg,~C.~J.; Kaupp,~M. Extended Benchmark Set of Main-Group Nuclear
  Shielding Constants and {NMR} Chemical Shifts and Its Use to Evaluate Modern
  {DFT} Methods. \emph{J. Chem. Theory Comput.} \textbf{2021}, \emph{17},
  7602--7621\relax
\mciteBstWouldAddEndPuncttrue
\mciteSetBstMidEndSepPunct{\mcitedefaultmidpunct}
{\mcitedefaultendpunct}{\mcitedefaultseppunct}\relax
\EndOfBibitem
\bibitem[Schattenberg \latin{et~al.}(2021)Schattenberg, Lehmann, Bühl, and
  Kaupp]{Schattenberg2021_JCTC_273}
Schattenberg,~C.~J.; Lehmann,~M.; Bühl,~M.; Kaupp,~M. Systematic Evaluation of
  Modern Density Functional Methods for the Computation of {NMR} Shifts of 3d
  Transition-Metal Nuclei. \emph{J. Chem. Theory Comput.} \textbf{2021},
  \emph{18}, 273--292\relax
\mciteBstWouldAddEndPuncttrue
\mciteSetBstMidEndSepPunct{\mcitedefaultmidpunct}
{\mcitedefaultendpunct}{\mcitedefaultseppunct}\relax
\EndOfBibitem
\bibitem[Bates and Furche(2012)Bates, and Furche]{Bates2012_JCP_164105}
Bates,~J.~E.; Furche,~F. {Harnessing the meta-generalized gradient
  approximation for time-dependent density functional theory.} \emph{J. Chem.
  Phys.} \textbf{2012}, \emph{137}, 164105\relax
\mciteBstWouldAddEndPuncttrue
\mciteSetBstMidEndSepPunct{\mcitedefaultmidpunct}
{\mcitedefaultendpunct}{\mcitedefaultseppunct}\relax
\EndOfBibitem
\bibitem[Bates \latin{et~al.}(2022)Bates, Heiche, Liang, and
  Furche]{Bates2022_JCP_159902}
Bates,~J.~E.; Heiche,~M.~C.; Liang,~J.; Furche,~F. Erratum: "{Harnessing} the
  meta-generalized gradient approximation for time-dependent density functional
  theory" [{J}. {Chem}. {Phys}. 137, 164105 (2012)]. \emph{J. Chem. Phys.}
  \textbf{2022}, \emph{156}, 159902\relax
\mciteBstWouldAddEndPuncttrue
\mciteSetBstMidEndSepPunct{\mcitedefaultmidpunct}
{\mcitedefaultendpunct}{\mcitedefaultseppunct}\relax
\EndOfBibitem
\bibitem[Grotjahn \latin{et~al.}(2022)Grotjahn, Furche, and
  Kaupp]{Grotjahn2022_JCP_111102}
Grotjahn,~R.; Furche,~F.; Kaupp,~M. Importance of imposing gauge invariance in
  time-dependent density functional theory calculations with meta-generalized
  gradient approximations. \emph{J. Chem. Phys.} \textbf{2022}, \emph{157},
  111102\relax
\mciteBstWouldAddEndPuncttrue
\mciteSetBstMidEndSepPunct{\mcitedefaultmidpunct}
{\mcitedefaultendpunct}{\mcitedefaultseppunct}\relax
\EndOfBibitem
\bibitem[Holzer \latin{et~al.}(2021)Holzer, Franzke, and
  Kehry]{Holzer2021_JCTC_2928}
Holzer,~C.; Franzke,~Y.~J.; Kehry,~M. Assessing the Accuracy of Local Hybrid
  Density Functional Approximations for Molecular Response Properties. \emph{J.
  Chem. Theory Comput.} \textbf{2021}, \emph{17}, 2928--2947\relax
\mciteBstWouldAddEndPuncttrue
\mciteSetBstMidEndSepPunct{\mcitedefaultmidpunct}
{\mcitedefaultendpunct}{\mcitedefaultseppunct}\relax
\EndOfBibitem
\bibitem[Franzke and Holzer(2022)Franzke, and Holzer]{Franzke2022_JCP_31102}
Franzke,~Y.~J.; Holzer,~C. Impact of the current density on paramagnetic {NMR}
  properties. \emph{J. Chem. Phys.} \textbf{2022}, \emph{157}, 031102\relax
\mciteBstWouldAddEndPuncttrue
\mciteSetBstMidEndSepPunct{\mcitedefaultmidpunct}
{\mcitedefaultendpunct}{\mcitedefaultseppunct}\relax
\EndOfBibitem
\bibitem[Liang \latin{et~al.}(2022)Liang, Feng, Hait, and
  Head-Gordon]{Liang2022_JCTC_3460}
Liang,~J.; Feng,~X.; Hait,~D.; Head-Gordon,~M. Revisiting the Performance of
  Time-Dependent Density Functional Theory for Electronic Excitations:
  Assessment of 43 Popular and Recently Developed Functionals from Rungs One to
  Four. \emph{J. Chem. Theory Comput.} \textbf{2022}, \emph{18},
  3460--3473\relax
\mciteBstWouldAddEndPuncttrue
\mciteSetBstMidEndSepPunct{\mcitedefaultmidpunct}
{\mcitedefaultendpunct}{\mcitedefaultseppunct}\relax
\EndOfBibitem
\bibitem[Neese(2022)]{Neese2022_WCMS_1606}
Neese,~F. Software update: The ORCA program system---Version 5.0. \emph{WIREs
  Comput. Mol. Sci.} \textbf{2022}, \emph{12}, e1606\relax
\mciteBstWouldAddEndPuncttrue
\mciteSetBstMidEndSepPunct{\mcitedefaultmidpunct}
{\mcitedefaultendpunct}{\mcitedefaultseppunct}\relax
\EndOfBibitem
\bibitem[Tellgren \latin{et~al.}(2014)Tellgren, Teale, Furness, Lange,
  Ekstr{\"{o}}m, and Helgaker]{Tellgren2014_JCP_34101}
Tellgren,~E.~I.; Teale,~A.~M.; Furness,~J.~W.; Lange,~K.~K.; Ekstr{\"{o}}m,~U.;
  Helgaker,~T. {Non-perturbative calculation of molecular magnetic properties
  within current-density functional theory}. \emph{J. Chem. Phys.}
  \textbf{2014}, \emph{140}, 034101\relax
\mciteBstWouldAddEndPuncttrue
\mciteSetBstMidEndSepPunct{\mcitedefaultmidpunct}
{\mcitedefaultendpunct}{\mcitedefaultseppunct}\relax
\EndOfBibitem
\bibitem[Furness \latin{et~al.}(2015)Furness, Verbeke, Tellgren, Stopkowicz,
  Ekstr{\"{o}}m, Helgaker, and Teale]{Furness2015_JCTC_4169}
Furness,~J.~W.; Verbeke,~J.; Tellgren,~E.~I.; Stopkowicz,~S.;
  Ekstr{\"{o}}m,~U.; Helgaker,~T.; Teale,~A.~M. {Current Density Functional
  Theory Using Meta-Generalized Gradient Exchange-Correlation Functionals}.
  \emph{J. Chem. Theory Comput.} \textbf{2015}, \emph{11}, 4169--4181\relax
\mciteBstWouldAddEndPuncttrue
\mciteSetBstMidEndSepPunct{\mcitedefaultmidpunct}
{\mcitedefaultendpunct}{\mcitedefaultseppunct}\relax
\EndOfBibitem
\bibitem[Reimann \latin{et~al.}(2015)Reimann, Ekstr{\"o}m, Stopkowicz, Teale,
  Borgoo, and Helgaker]{Reimann2015_PCCP_18834}
Reimann,~S.; Ekstr{\"o}m,~U.; Stopkowicz,~S.; Teale,~A.~M.; Borgoo,~A.;
  Helgaker,~T. The importance of current contributions to shielding constants
  in density-functional theory. \emph{Phys. Chem. Chem. Phys.} \textbf{2015},
  \emph{17}, 18834--18842\relax
\mciteBstWouldAddEndPuncttrue
\mciteSetBstMidEndSepPunct{\mcitedefaultmidpunct}
{\mcitedefaultendpunct}{\mcitedefaultseppunct}\relax
\EndOfBibitem
\bibitem[Irons \latin{et~al.}(2020)Irons, Spence, David, Speake, Helgaker, and
  Teale]{Irons2020_JPCA_1321}
Irons,~T. J.~P.; Spence,~L.; David,~G.; Speake,~B.~T.; Helgaker,~T.;
  Teale,~A.~M. {Analyzing Magnetically Induced Currents in Molecular Systems
  Using Current-Density-Functional Theory}. \emph{J. Phys. Chem. A}
  \textbf{2020}, \emph{124}, 1321--1333\relax
\mciteBstWouldAddEndPuncttrue
\mciteSetBstMidEndSepPunct{\mcitedefaultmidpunct}
{\mcitedefaultendpunct}{\mcitedefaultseppunct}\relax
\EndOfBibitem
\bibitem[Irons \latin{et~al.}(2021)Irons, David, and
  Teale]{Irons2021_JCTC_2166}
Irons,~T. J.~P.; David,~G.; Teale,~A.~M. Optimizing Molecular Geometries in
  Strong Magnetic Fields. \emph{J. Chem. Theory Comput.} \textbf{2021},
  \emph{17}, 2166--2185\relax
\mciteBstWouldAddEndPuncttrue
\mciteSetBstMidEndSepPunct{\mcitedefaultmidpunct}
{\mcitedefaultendpunct}{\mcitedefaultseppunct}\relax
\EndOfBibitem
\bibitem[Jaramillo \latin{et~al.}(2003)Jaramillo, Scuseria, and
  Ernzerhof]{Jaramillo2003_JCP_1068}
Jaramillo,~J.; Scuseria,~G.~E.; Ernzerhof,~M. {Local hybrid functionals}.
  \emph{J. Chem. Phys.} \textbf{2003}, \emph{118}, 1068--1073\relax
\mciteBstWouldAddEndPuncttrue
\mciteSetBstMidEndSepPunct{\mcitedefaultmidpunct}
{\mcitedefaultendpunct}{\mcitedefaultseppunct}\relax
\EndOfBibitem
\bibitem[Maier \latin{et~al.}(2019)Maier, Arbuznikov, and
  Kaupp]{Maier2019_WIRCMS_1378}
Maier,~T.~M.; Arbuznikov,~A.~V.; Kaupp,~M. {Local hybrid functionals: Theory,
  implementation, and performance of an emerging new tool in quantum chemistry
  and beyond}. \emph{Wiley Interdiscip. Rev. Comput. Mol. Sci.} \textbf{2019},
  \emph{9}, e1378\relax
\mciteBstWouldAddEndPuncttrue
\mciteSetBstMidEndSepPunct{\mcitedefaultmidpunct}
{\mcitedefaultendpunct}{\mcitedefaultseppunct}\relax
\EndOfBibitem
\bibitem[Weymuth and Reiher(2022)Weymuth, and Reiher]{Weymuth2022_PCCP_14692}
Weymuth,~T.; Reiher,~M. The transferability limits of static benchmarks.
  \emph{Phys. Chem. Chem. Phys.} \textbf{2022}, \emph{24}, 14692--14698\relax
\mciteBstWouldAddEndPuncttrue
\mciteSetBstMidEndSepPunct{\mcitedefaultmidpunct}
{\mcitedefaultendpunct}{\mcitedefaultseppunct}\relax
\EndOfBibitem
\bibitem[Fürst \latin{et~al.}(2023)Fürst, Haasler, Grotjahn, and
  Kaupp]{Fuerst2023_JCTC_488}
Fürst,~S.; Haasler,~M.; Grotjahn,~R.; Kaupp,~M. Full Implementation,
  Optimization, and Evaluation of a Range-Separated Local Hybrid Functional
  with Wide Accuracy for Ground and Excited States. \emph{J. Chem. Theory
  Comput.} \textbf{2023}, \emph{19}, 488--502\relax
\mciteBstWouldAddEndPuncttrue
\mciteSetBstMidEndSepPunct{\mcitedefaultmidpunct}
{\mcitedefaultendpunct}{\mcitedefaultseppunct}\relax
\EndOfBibitem
\bibitem[Fürst and Kaupp(2023)Fürst, and Kaupp]{Fuerst2023_JCTC_sub}
Fürst,~S.; Kaupp,~M. Accurate ionization potentials, electron affinities and
  band gaps from the $\omega$LH22t range-separated local hybrid functional: no
  tuning required. \emph{J. Chem. Theory Comput.} \textbf{2023}, \relax
\mciteBstWouldAddEndPunctfalse
\mciteSetBstMidEndSepPunct{\mcitedefaultmidpunct}
{}{\mcitedefaultseppunct}\relax
\EndOfBibitem
\bibitem[Wody\'{n}ski and Kaupp(2022)Wody\'{n}ski, and
  Kaupp]{Wodynski2022_JCTC_6111}
Wody\'{n}ski,~A.; Kaupp,~M. Local Hybrid Functional Applicable to Weakly and
  Strongly Correlated Systems. \emph{J. Chem. Theory Comput.} \textbf{2022},
  \emph{18}, 6111--6123\relax
\mciteBstWouldAddEndPuncttrue
\mciteSetBstMidEndSepPunct{\mcitedefaultmidpunct}
{\mcitedefaultendpunct}{\mcitedefaultseppunct}\relax
\EndOfBibitem
\bibitem[Wody{\'{n}}ski \latin{et~al.}(2023)Wody{\'{n}}ski, Arbuznikov, and
  Kaupp]{Wodynski2023_JCP_244117}
Wody{\'{n}}ski,~A.; Arbuznikov,~A.~V.; Kaupp,~M. Strong-correlation density
  functionals made simple. \emph{J. Chem. Phys.} \textbf{2023}, \emph{158},
  244117\relax
\mciteBstWouldAddEndPuncttrue
\mciteSetBstMidEndSepPunct{\mcitedefaultmidpunct}
{\mcitedefaultendpunct}{\mcitedefaultseppunct}\relax
\EndOfBibitem
\bibitem[Ruud \latin{et~al.}(1993)Ruud, Helgaker, Bak, J{\o}rgensen, and
  Jensen]{Ruud1993_JCP_3847}
Ruud,~K.; Helgaker,~T.; Bak,~K.~L.; J{\o}rgensen,~P.; Jensen,~H. J.~A.
  {Hartree--Fock limit magnetizabilities from London orbitals}. \emph{J. Chem.
  Phys.} \textbf{1993}, \emph{99}, 3847\relax
\mciteBstWouldAddEndPuncttrue
\mciteSetBstMidEndSepPunct{\mcitedefaultmidpunct}
{\mcitedefaultendpunct}{\mcitedefaultseppunct}\relax
\EndOfBibitem
\bibitem[Ruud \latin{et~al.}(1994)Ruud, Skaane, Helgaker, Bak, and
  J\o{}rgensen]{Ruud1994_JACS_10135}
Ruud,~K.; Skaane,~H.; Helgaker,~T.; Bak,~K.~L.; J\o{}rgensen,~P.
  Magnetizability of Hydrocarbons. \emph{J. Am. Chem. Soc.} \textbf{1994},
  \emph{116}, 10135--10140\relax
\mciteBstWouldAddEndPuncttrue
\mciteSetBstMidEndSepPunct{\mcitedefaultmidpunct}
{\mcitedefaultendpunct}{\mcitedefaultseppunct}\relax
\EndOfBibitem
\bibitem[Ruud \latin{et~al.}(1995)Ruud, Helgaker, Bak, J{\o}rgensen, and
  Olsen]{Ruud1995_CP_157}
Ruud,~K.; Helgaker,~T.; Bak,~K.~L.; J{\o}rgensen,~P.; Olsen,~J. Accurate
  magnetizabilities of the isoelectronic series \ce{BeH-}, \ce{BH}, and
  \ce{CH+}. The {MCSCF}-{GIAO} approach. \emph{Chem. Phys.} \textbf{1995},
  \emph{195}, 157--169\relax
\mciteBstWouldAddEndPuncttrue
\mciteSetBstMidEndSepPunct{\mcitedefaultmidpunct}
{\mcitedefaultendpunct}{\mcitedefaultseppunct}\relax
\EndOfBibitem
\bibitem[Helgaker \latin{et~al.}(2012)Helgaker, Coriani, J{\o}rgensen,
  Kristensen, Olsen, and Ruud]{Helgaker2012_CR_543}
Helgaker,~T.; Coriani,~S.; J{\o}rgensen,~P.; Kristensen,~K.; Olsen,~J.;
  Ruud,~K. {Recent advances in wave function-based methods of
  molecular-property calculations}. \emph{Chem. Rev.} \textbf{2012},
  \emph{112}, 543--631\relax
\mciteBstWouldAddEndPuncttrue
\mciteSetBstMidEndSepPunct{\mcitedefaultmidpunct}
{\mcitedefaultendpunct}{\mcitedefaultseppunct}\relax
\EndOfBibitem
\bibitem[Stevens \latin{et~al.}(1963)Stevens, Pitzer, and
  Lipscomb]{Stevens1963_JCP_550}
Stevens,~R.~M.; Pitzer,~R.~M.; Lipscomb,~W.~N. Perturbed {Hartree}--{Fock}
  Calculations. {I}. Magnetic Susceptibility and Shielding in the {LiH}
  Molecule. \emph{J. Chem. Phys.} \textbf{1963}, \emph{38}, 550--560\relax
\mciteBstWouldAddEndPuncttrue
\mciteSetBstMidEndSepPunct{\mcitedefaultmidpunct}
{\mcitedefaultendpunct}{\mcitedefaultseppunct}\relax
\EndOfBibitem
\bibitem[Jameson and Buckingham(1979)Jameson, and
  Buckingham]{Jameson1979_JPC_3366}
Jameson,~C.~J.; Buckingham,~A.~D. Nuclear magnetic shielding density. \emph{J.
  Phys. Chem.} \textbf{1979}, \emph{83}, 3366--3371\relax
\mciteBstWouldAddEndPuncttrue
\mciteSetBstMidEndSepPunct{\mcitedefaultmidpunct}
{\mcitedefaultendpunct}{\mcitedefaultseppunct}\relax
\EndOfBibitem
\bibitem[Jameson and Buckingham(1980)Jameson, and
  Buckingham]{Jameson1980_JCP_5684}
Jameson,~C.~J.; Buckingham,~A.~D. Molecular electronic property density
  functions: The nuclear magnetic shielding density. \emph{J. Chem. Phys.}
  \textbf{1980}, \emph{73}, 5684--5692\relax
\mciteBstWouldAddEndPuncttrue
\mciteSetBstMidEndSepPunct{\mcitedefaultmidpunct}
{\mcitedefaultendpunct}{\mcitedefaultseppunct}\relax
\EndOfBibitem
\bibitem[Fowler \latin{et~al.}(1998)Fowler, Steiner, Cadioli, and
  Zanasi]{Fowler1998_JPCA_7297}
Fowler,~P.~W.; Steiner,~E.; Cadioli,~B.; Zanasi,~R. Distributed-gauge
  calculations of current density maps, magnetizabilities, and shieldings for a
  series of neutral and dianionic fused tetracycles: pyracylene (\ce{C14H8}),
  acepleiadylene (\ce{C16H10}), and dipleiadiene (\ce{C18H12}). \emph{J. Phys.
  Chem. A} \textbf{1998}, \emph{102}, 7297--7302\relax
\mciteBstWouldAddEndPuncttrue
\mciteSetBstMidEndSepPunct{\mcitedefaultmidpunct}
{\mcitedefaultendpunct}{\mcitedefaultseppunct}\relax
\EndOfBibitem
\bibitem[Ilia{\v{s}} \latin{et~al.}(2013)Ilia{\v{s}}, Jensen, Bast, and
  Saue]{Ilias2013_MP_1373}
Ilia{\v{s}},~M.; Jensen,~H. J.~A.; Bast,~R.; Saue,~T. Gauge origin independent
  calculations of molecular magnetisabilities in relativistic four-component
  theory. \emph{Mol. Phys.} \textbf{2013}, \emph{111}, 1373--1381\relax
\mciteBstWouldAddEndPuncttrue
\mciteSetBstMidEndSepPunct{\mcitedefaultmidpunct}
{\mcitedefaultendpunct}{\mcitedefaultseppunct}\relax
\EndOfBibitem
\bibitem[Lazzeretti(2000)]{Lazzeretti2000_PNMRS_1}
Lazzeretti,~P. Ring currents. \emph{Prog. Nucl. Magn. Reson. Spectrosc.}
  \textbf{2000}, \emph{36}, 1--88\relax
\mciteBstWouldAddEndPuncttrue
\mciteSetBstMidEndSepPunct{\mcitedefaultmidpunct}
{\mcitedefaultendpunct}{\mcitedefaultseppunct}\relax
\EndOfBibitem
\bibitem[Lazzeretti(2018)]{Lazzeretti2018_JCP_134109}
Lazzeretti,~P. Current density tensors. \emph{J. Chem. Phys.} \textbf{2018},
  \emph{148}, 134109\relax
\mciteBstWouldAddEndPuncttrue
\mciteSetBstMidEndSepPunct{\mcitedefaultmidpunct}
{\mcitedefaultendpunct}{\mcitedefaultseppunct}\relax
\EndOfBibitem
\bibitem[Sambe(1973)]{Sambe1973_JCP_555}
Sambe,~H. Properties of induced electron current density of a molecule under a
  static uniform magnetic field. \emph{J. Chem. Phys.} \textbf{1973},
  \emph{59}, 555--555\relax
\mciteBstWouldAddEndPuncttrue
\mciteSetBstMidEndSepPunct{\mcitedefaultmidpunct}
{\mcitedefaultendpunct}{\mcitedefaultseppunct}\relax
\EndOfBibitem
\bibitem[Jus{\'{e}}lius \latin{et~al.}(2004)Jus{\'{e}}lius, Sundholm, and
  Gauss]{Juselius2004_JCP_3952}
Jus{\'{e}}lius,~J.; Sundholm,~D.; Gauss,~J. Calculation of current densities
  using gauge-including atomic orbitals. \emph{J. Chem. Phys.} \textbf{2004},
  \emph{121}, 3952--3963\relax
\mciteBstWouldAddEndPuncttrue
\mciteSetBstMidEndSepPunct{\mcitedefaultmidpunct}
{\mcitedefaultendpunct}{\mcitedefaultseppunct}\relax
\EndOfBibitem
\bibitem[Taubert \latin{et~al.}(2011)Taubert, Sundholm, and
  Jus{\'{e}}lius]{Taubert2011_JCP_54123}
Taubert,~S.; Sundholm,~D.; Jus{\'{e}}lius,~J. Calculation of spin-current
  densities using gauge-including atomic orbitals. \emph{J. Chem. Phys.}
  \textbf{2011}, \emph{134}, 054123\relax
\mciteBstWouldAddEndPuncttrue
\mciteSetBstMidEndSepPunct{\mcitedefaultmidpunct}
{\mcitedefaultendpunct}{\mcitedefaultseppunct}\relax
\EndOfBibitem
\bibitem[Fliegl \latin{et~al.}(2011)Fliegl, Taubert, Lehtonen, and
  Sundholm]{Fliegl2011_PCCP_20500}
Fliegl,~H.; Taubert,~S.; Lehtonen,~O.; Sundholm,~D. The gauge including
  magnetically induced current method. \emph{Phys. Chem. Chem. Phys.}
  \textbf{2011}, \emph{13}, 20500\relax
\mciteBstWouldAddEndPuncttrue
\mciteSetBstMidEndSepPunct{\mcitedefaultmidpunct}
{\mcitedefaultendpunct}{\mcitedefaultseppunct}\relax
\EndOfBibitem
\bibitem[Sundholm \latin{et~al.}(2016)Sundholm, Fliegl, and
  Berger]{Sundholm2016_WIRCMS_639}
Sundholm,~D.; Fliegl,~H.; Berger,~R.~J. Calculations of magnetically induced
  current densities: theory and applications. \emph{Wiley Interdiscip. Rev.:
  Comput. Mol. Sci.} \textbf{2016}, \emph{6}, 639--678\relax
\mciteBstWouldAddEndPuncttrue
\mciteSetBstMidEndSepPunct{\mcitedefaultmidpunct}
{\mcitedefaultendpunct}{\mcitedefaultseppunct}\relax
\EndOfBibitem
\bibitem[Steiner and Fowler(2004)Steiner, and Fowler]{Steiner2004_PCCP_261}
Steiner,~E.; Fowler,~P.~W. On the orbital analysis of magnetic properties.
  \emph{Phys. Chem. Chem. Phys.} \textbf{2004}, \emph{6}, 261--272\relax
\mciteBstWouldAddEndPuncttrue
\mciteSetBstMidEndSepPunct{\mcitedefaultmidpunct}
{\mcitedefaultendpunct}{\mcitedefaultseppunct}\relax
\EndOfBibitem
\bibitem[Pelloni \latin{et~al.}(2004)Pelloni, Ligabue, and
  Lazzeretti]{Pelloni2004_OL_4451}
Pelloni,~S.; Ligabue,~A.; Lazzeretti,~P. Ring-current models from the
  differential {Biot--Savart} law. \emph{Org. Lett.} \textbf{2004}, \emph{6},
  4451--4454\relax
\mciteBstWouldAddEndPuncttrue
\mciteSetBstMidEndSepPunct{\mcitedefaultmidpunct}
{\mcitedefaultendpunct}{\mcitedefaultseppunct}\relax
\EndOfBibitem
\bibitem[Ferraro \latin{et~al.}(2004)Ferraro, Lazzeretti, Viglione, and
  Zanasi]{Ferraro2004_CPL_268}
Ferraro,~M.~B.; Lazzeretti,~P.; Viglione,~R.~G.; Zanasi,~R. Understanding
  proton magnetic shielding in the benzene molecule. \emph{Chem. Phys. Lett.}
  \textbf{2004}, \emph{390}, 268--271\relax
\mciteBstWouldAddEndPuncttrue
\mciteSetBstMidEndSepPunct{\mcitedefaultmidpunct}
{\mcitedefaultendpunct}{\mcitedefaultseppunct}\relax
\EndOfBibitem
\bibitem[Soncini \latin{et~al.}(2005)Soncini, Fowler, Lazzeretti, and
  Zanasi]{Soncini2005_CPL_164}
Soncini,~A.; Fowler,~P.; Lazzeretti,~P.; Zanasi,~R. {Ring-current signatures in
  shielding-density maps}. \emph{Chem. Phys. Lett.} \textbf{2005}, \emph{401},
  164--169\relax
\mciteBstWouldAddEndPuncttrue
\mciteSetBstMidEndSepPunct{\mcitedefaultmidpunct}
{\mcitedefaultendpunct}{\mcitedefaultseppunct}\relax
\EndOfBibitem
\bibitem[Ferraro \latin{et~al.}(2005)Ferraro, Faglioni, Ligabue, Pelloni, and
  Lazzeretti]{Ferraro2005_MRC_316}
Ferraro,~M.~B.; Faglioni,~F.; Ligabue,~A.; Pelloni,~S.; Lazzeretti,~P. Ring
  current effects on nuclear magnetic shielding of carbon in the benzene
  molecule. \emph{Magn. Reson. Chem.} \textbf{2005}, \emph{43}, 316--320\relax
\mciteBstWouldAddEndPuncttrue
\mciteSetBstMidEndSepPunct{\mcitedefaultmidpunct}
{\mcitedefaultendpunct}{\mcitedefaultseppunct}\relax
\EndOfBibitem
\bibitem[Acke \latin{et~al.}(2018)Acke, {Van Damme}, Havenith, and
  Bultinck]{Acke2018_JCC_511}
Acke,~G.; {Van Damme},~S.; Havenith,~R. W.~A.; Bultinck,~P. Interpreting the
  behavior of the NICS$_{zz}$ by resolving in orbitals, sign, and positions.
  \emph{J. Comput. Chem.} \textbf{2018}, \emph{39}, 511--519\relax
\mciteBstWouldAddEndPuncttrue
\mciteSetBstMidEndSepPunct{\mcitedefaultmidpunct}
{\mcitedefaultendpunct}{\mcitedefaultseppunct}\relax
\EndOfBibitem
\bibitem[Acke \latin{et~al.}(2019)Acke, {Van Damme}, Havenith, and
  Bultinck]{Acke2019_PCCP_3145}
Acke,~G.; {Van Damme},~S.; Havenith,~R. W.~A.; Bultinck,~P. {Quantifying the
  conceptual problems associated with the isotropic NICS through analyses of
  its underlying density}. \emph{Phys. Chem. Chem. Phys.} \textbf{2019},
  \emph{21}, 3145--3153\relax
\mciteBstWouldAddEndPuncttrue
\mciteSetBstMidEndSepPunct{\mcitedefaultmidpunct}
{\mcitedefaultendpunct}{\mcitedefaultseppunct}\relax
\EndOfBibitem
\bibitem[Jinger \latin{et~al.}(2021)Jinger, Fliegl, Bast, Dimitrova, Lehtola,
  and Sundholm]{Jinger2021_JPCA_1778}
Jinger,~R.~K.; Fliegl,~H.; Bast,~R.; Dimitrova,~M.; Lehtola,~S.; Sundholm,~D.
  Spatial Contributions to Nuclear Magnetic Shieldings. \emph{J. Phys. Chem. A}
  \textbf{2021}, \emph{125}, 1778--1786\relax
\mciteBstWouldAddEndPuncttrue
\mciteSetBstMidEndSepPunct{\mcitedefaultmidpunct}
{\mcitedefaultendpunct}{\mcitedefaultseppunct}\relax
\EndOfBibitem
\bibitem[Fliegl \latin{et~al.}(2021)Fliegl, Dimitrova, Berger, and
  Sundholm]{Fliegl2021_C_1005}
Fliegl,~H.; Dimitrova,~M.; Berger,~R. J.~F.; Sundholm,~D. Spatial Contributions
  to $^1$H {NMR} Chemical Shifts of Free-Base Porphyrinoids. \emph{Chemistry}
  \textbf{2021}, \emph{3}, 1005--1021\relax
\mciteBstWouldAddEndPuncttrue
\mciteSetBstMidEndSepPunct{\mcitedefaultmidpunct}
{\mcitedefaultendpunct}{\mcitedefaultseppunct}\relax
\EndOfBibitem
\bibitem[Summa \latin{et~al.}(2021)Summa, Monaco, and Zanasi]{Summa2021_CCA_43}
Summa,~F.~F.; Monaco,~G.; Zanasi,~R. Decomposition of Molecular Integrals into
  Atomic Contributions via Becke Partitioning Scheme: a Caveat. \emph{Croat.
  Chem. Acta} \textbf{2021}, \emph{94}, 43--46\relax
\mciteBstWouldAddEndPuncttrue
\mciteSetBstMidEndSepPunct{\mcitedefaultmidpunct}
{\mcitedefaultendpunct}{\mcitedefaultseppunct}\relax
\EndOfBibitem
\bibitem[Sundholm \latin{et~al.}(2021)Sundholm, Dimitrova, and
  Berger]{Sundholm2021_CC_12362}
Sundholm,~D.; Dimitrova,~M.; Berger,~R. J.~F. Current density and molecular
  magnetic properties. \emph{Chem. Commun.} \textbf{2021}, \emph{57},
  12362--12378\relax
\mciteBstWouldAddEndPuncttrue
\mciteSetBstMidEndSepPunct{\mcitedefaultmidpunct}
{\mcitedefaultendpunct}{\mcitedefaultseppunct}\relax
\EndOfBibitem
\bibitem[Hameka(1962)]{Hameka1962_RMP_87}
Hameka,~H.~F. Theory of Magnetic Properties of Molecules with Particular
  Emphasis on the Hydrogen Molecule. \emph{Rev. Mod. Phys.} \textbf{1962},
  \emph{34}, 87--101\relax
\mciteBstWouldAddEndPuncttrue
\mciteSetBstMidEndSepPunct{\mcitedefaultmidpunct}
{\mcitedefaultendpunct}{\mcitedefaultseppunct}\relax
\EndOfBibitem
\bibitem[Ditchfield(1974)]{Ditchfield1974_MP_789}
Ditchfield,~R. {Self-consistent perturbation theory of diamagnetism. I. A
  gauge-invariant LCAO method for N.M.R. chemical shifts}. \emph{Mol. Phys.}
  \textbf{1974}, \emph{27}, 789--807\relax
\mciteBstWouldAddEndPuncttrue
\mciteSetBstMidEndSepPunct{\mcitedefaultmidpunct}
{\mcitedefaultendpunct}{\mcitedefaultseppunct}\relax
\EndOfBibitem
\bibitem[Wolinski \latin{et~al.}(1990)Wolinski, Hinton, and
  Pulay]{Wolinski1990_JACS_8251}
Wolinski,~K.; Hinton,~J.~F.; Pulay,~P. Efficient implementation of the
  gauge-independent atomic orbital method for {NMR} chemical shift
  calculations. \emph{J. Am. Chem. Soc.} \textbf{1990}, \emph{112},
  8251--8260\relax
\mciteBstWouldAddEndPuncttrue
\mciteSetBstMidEndSepPunct{\mcitedefaultmidpunct}
{\mcitedefaultendpunct}{\mcitedefaultseppunct}\relax
\EndOfBibitem
\bibitem[Gauss(1992)]{Gauss1992_CPL_614}
Gauss,~J. Calculation of {NMR} chemical shifts at second-order many-body
  perturbation theory using gauge-including atomic orbitals. \emph{Chem. Phys.
  Lett.} \textbf{1992}, \emph{191}, 614--620\relax
\mciteBstWouldAddEndPuncttrue
\mciteSetBstMidEndSepPunct{\mcitedefaultmidpunct}
{\mcitedefaultendpunct}{\mcitedefaultseppunct}\relax
\EndOfBibitem
\bibitem[London(1937)]{London1937_JPlR_397}
London,~F. {Th{\'{e}}orie quantique des courants interatomiques dans les
  combinaisons aromatiques}. \emph{J. Phys. le Radium} \textbf{1937}, \emph{8},
  397--409\relax
\mciteBstWouldAddEndPuncttrue
\mciteSetBstMidEndSepPunct{\mcitedefaultmidpunct}
{\mcitedefaultendpunct}{\mcitedefaultseppunct}\relax
\EndOfBibitem
\bibitem[Vignale and Rasolt(1987)Vignale, and Rasolt]{Vignale1987_PRL_2360}
Vignale,~G.; Rasolt,~M. {Density-functional theory in strong magnetic fields}.
  \emph{Phys. Rev. Lett.} \textbf{1987}, \emph{59}, 2360--2363\relax
\mciteBstWouldAddEndPuncttrue
\mciteSetBstMidEndSepPunct{\mcitedefaultmidpunct}
{\mcitedefaultendpunct}{\mcitedefaultseppunct}\relax
\EndOfBibitem
\bibitem[Vignale and Rasolt(1988)Vignale, and Rasolt]{Vignale1988_PRB_10685}
Vignale,~G.; Rasolt,~M. Current- and spin-density-functional theory for
  inhomogeneous electronic systems in strong magnetic fields. \emph{Phys. Rev.
  B} \textbf{1988}, \emph{37}, 10685--10696\relax
\mciteBstWouldAddEndPuncttrue
\mciteSetBstMidEndSepPunct{\mcitedefaultmidpunct}
{\mcitedefaultendpunct}{\mcitedefaultseppunct}\relax
\EndOfBibitem
\bibitem[Sagvolden \latin{et~al.}(2013)Sagvolden, Ekstrm, and
  Tellgren]{Sagvolden2013_MP_1295}
Sagvolden,~E.; Ekstrm,~U.; Tellgren,~E.~I. Isoorbital indicators for current
  density functional theory. \emph{Mol. Phys.} \textbf{2013}, \emph{111},
  1295--1302\relax
\mciteBstWouldAddEndPuncttrue
\mciteSetBstMidEndSepPunct{\mcitedefaultmidpunct}
{\mcitedefaultendpunct}{\mcitedefaultseppunct}\relax
\EndOfBibitem
\bibitem[Tao(2005)]{Tao2005_PRB_205107}
Tao,~J. Explicit inclusion of paramagnetic current density in the
  exchange-correlation functionals of current-density functional theory.
  \emph{Phys. Rev. B} \textbf{2005}, \emph{71}, 205107\relax
\mciteBstWouldAddEndPuncttrue
\mciteSetBstMidEndSepPunct{\mcitedefaultmidpunct}
{\mcitedefaultendpunct}{\mcitedefaultseppunct}\relax
\EndOfBibitem
\bibitem[Dobson(1992)]{Dobson1992_JPCM_7877}
Dobson,~J.~F. Spin-density functionals for the electron correlation energy with
  automatic freedom from orbital self-interaction. \emph{J. Phys.: Condens.
  Matter} \textbf{1992}, \emph{4}, 7877--7890\relax
\mciteBstWouldAddEndPuncttrue
\mciteSetBstMidEndSepPunct{\mcitedefaultmidpunct}
{\mcitedefaultendpunct}{\mcitedefaultseppunct}\relax
\EndOfBibitem
\bibitem[von Weizs\"acker(1935)]{Weizsacker1935_431}
von Weizs\"acker,~C.~F. Zur {Theorie} der {Kernmassen}. \emph{Z. Phys.}
  \textbf{1935}, \emph{96}, 431\relax
\mciteBstWouldAddEndPuncttrue
\mciteSetBstMidEndSepPunct{\mcitedefaultmidpunct}
{\mcitedefaultendpunct}{\mcitedefaultseppunct}\relax
\EndOfBibitem
\bibitem[Burke \latin{et~al.}(1998)Burke, Cruz, and Lam]{Burke1998_JCP_8161}
Burke,~K.; Cruz,~F.~G.; Lam,~K.-C. Unambiguous exchange-correlation energy
  density. \emph{J. Chem. Phys.} \textbf{1998}, \emph{109}, 8161--8167\relax
\mciteBstWouldAddEndPuncttrue
\mciteSetBstMidEndSepPunct{\mcitedefaultmidpunct}
{\mcitedefaultendpunct}{\mcitedefaultseppunct}\relax
\EndOfBibitem
\bibitem[Cruz \latin{et~al.}(1998)Cruz, Lam, and Burke]{Cruz1998_JPCA_4911}
Cruz,~F.~G.; Lam,~K.-C.; Burke,~K. Exchange-Correlation Energy Density from
  Virial Theorem. \emph{J. Phys. Chem. A} \textbf{1998}, \emph{102},
  4911--4917\relax
\mciteBstWouldAddEndPuncttrue
\mciteSetBstMidEndSepPunct{\mcitedefaultmidpunct}
{\mcitedefaultendpunct}{\mcitedefaultseppunct}\relax
\EndOfBibitem
\bibitem[Maier \latin{et~al.}(2016)Maier, Haasler, Arbuznikov, and
  Kaupp]{Maier2016_PCCP_21133}
Maier,~T.~M.; Haasler,~M.; Arbuznikov,~A.~V.; Kaupp,~M. {New approaches for the
  calibration of exchange-energy densities in local hybrid functionals}.
  \emph{Phys. Chem. Chem. Phys.} \textbf{2016}, \emph{18}, 21133--21144\relax
\mciteBstWouldAddEndPuncttrue
\mciteSetBstMidEndSepPunct{\mcitedefaultmidpunct}
{\mcitedefaultendpunct}{\mcitedefaultseppunct}\relax
\EndOfBibitem
\bibitem[Tao \latin{et~al.}(2008)Tao, Staroverov, Scuseria, and
  Perdew]{Tao2008_PRA_012509}
Tao,~J.; Staroverov,~V.~N.; Scuseria,~G.~E.; Perdew,~J.~P. Exact-exchange
  energy density in the gauge of a semilocal density-functional approximation.
  \emph{Phys. Rev. A} \textbf{2008}, \emph{77}, 012509\relax
\mciteBstWouldAddEndPuncttrue
\mciteSetBstMidEndSepPunct{\mcitedefaultmidpunct}
{\mcitedefaultendpunct}{\mcitedefaultseppunct}\relax
\EndOfBibitem
\bibitem[Haasler \latin{et~al.}(2020)Haasler, Maier, Grotjahn, G{\"{u}}ckel,
  Arbuznikov, and Kaupp]{Haasler2020_JCTC_5645}
Haasler,~M.; Maier,~T.~M.; Grotjahn,~R.; G{\"{u}}ckel,~S.; Arbuznikov,~A.~V.;
  Kaupp,~M. {A Local Hybrid Functional with Wide Applicability and Good Balance
  between (De)Localization and Left--Right Correlation}. \emph{J. Chem. Theory
  Comput.} \textbf{2020}, \emph{16}, 5645--5657\relax
\mciteBstWouldAddEndPuncttrue
\mciteSetBstMidEndSepPunct{\mcitedefaultmidpunct}
{\mcitedefaultendpunct}{\mcitedefaultseppunct}\relax
\EndOfBibitem
\bibitem[Wody{\'{n}}ski \latin{et~al.}(2021)Wody{\'{n}}ski, Arbuznikov, and
  Kaupp]{Wodynski2021_JCP_144101}
Wody{\'{n}}ski,~A.; Arbuznikov,~A.~V.; Kaupp,~M. Local hybrid functionals
  augmented by a strong-correlation model. \emph{J. Chem. Phys.} \textbf{2021},
  \emph{155}, 144101\relax
\mciteBstWouldAddEndPuncttrue
\mciteSetBstMidEndSepPunct{\mcitedefaultmidpunct}
{\mcitedefaultendpunct}{\mcitedefaultseppunct}\relax
\EndOfBibitem
\bibitem[Kong and Proynov(2016)Kong, and Proynov]{Kong2016_JCTC_133}
Kong,~J.; Proynov,~E. {Density Functional Model for Nondynamic and Strong
  Correlation}. \emph{J. Chem. Theory Comput.} \textbf{2016}, \emph{12},
  133--143\relax
\mciteBstWouldAddEndPuncttrue
\mciteSetBstMidEndSepPunct{\mcitedefaultmidpunct}
{\mcitedefaultendpunct}{\mcitedefaultseppunct}\relax
\EndOfBibitem
\bibitem[Becke(2013)]{Becke2013_JCP_74109}
Becke,~A.~D. {Density functionals for static, dynamical, and strong
  correlation}. \emph{J. Chem. Phys.} \textbf{2013}, \emph{138}, 074109\relax
\mciteBstWouldAddEndPuncttrue
\mciteSetBstMidEndSepPunct{\mcitedefaultmidpunct}
{\mcitedefaultendpunct}{\mcitedefaultseppunct}\relax
\EndOfBibitem
\bibitem[Lehtola \latin{et~al.}(2018)Lehtola, Steigemann, Oliveira, and
  Marques]{Lehtola2018_S_1}
Lehtola,~S.; Steigemann,~C.; Oliveira,~M. J.~T.; Marques,~M. A.~L. Recent
  developments in {LIBXC}---a comprehensive library of functionals for density
  functional theory. \emph{SoftwareX} \textbf{2018}, \emph{7}, 1--5\relax
\mciteBstWouldAddEndPuncttrue
\mciteSetBstMidEndSepPunct{\mcitedefaultmidpunct}
{\mcitedefaultendpunct}{\mcitedefaultseppunct}\relax
\EndOfBibitem
\bibitem[Dunning(1989)]{Dunning1989_JCP_1007}
Dunning,~T.~H. {Gaussian basis sets for use in correlated molecular
  calculations. I. The atoms boron through neon and hydrogen}. \emph{J. Chem.
  Phys.} \textbf{1989}, \emph{90}, 1007\relax
\mciteBstWouldAddEndPuncttrue
\mciteSetBstMidEndSepPunct{\mcitedefaultmidpunct}
{\mcitedefaultendpunct}{\mcitedefaultseppunct}\relax
\EndOfBibitem
\bibitem[Kendall \latin{et~al.}(1992)Kendall, Dunning, and
  Harrison]{Kendall1992_JCP_6796}
Kendall,~R.~A.; Dunning,~T.~H.; Harrison,~R.~J. {Electron affinities of the
  first-row atoms revisited. Systematic basis sets and wave functions}.
  \emph{J. Chem. Phys.} \textbf{1992}, \emph{96}, 6796\relax
\mciteBstWouldAddEndPuncttrue
\mciteSetBstMidEndSepPunct{\mcitedefaultmidpunct}
{\mcitedefaultendpunct}{\mcitedefaultseppunct}\relax
\EndOfBibitem
\bibitem[Woon and Dunning(1993)Woon, and Dunning]{Woon1993_JCP_1358}
Woon,~D.~E.; Dunning,~T.~H. {Gaussian basis sets for use in correlated
  molecular calculations. III. The atoms aluminum through argon}. \emph{J.
  Chem. Phys.} \textbf{1993}, \emph{98}, 1358\relax
\mciteBstWouldAddEndPuncttrue
\mciteSetBstMidEndSepPunct{\mcitedefaultmidpunct}
{\mcitedefaultendpunct}{\mcitedefaultseppunct}\relax
\EndOfBibitem
\bibitem[Woon and Dunning(1995)Woon, and Dunning]{Woon1995_JCP_4572}
Woon,~D.~E.; Dunning,~T.~H. {Gaussian basis sets for use in correlated
  molecular calculations. V. Core-valence basis sets for boron through neon}.
  \emph{J. Chem. Phys.} \textbf{1995}, \emph{103}, 4572\relax
\mciteBstWouldAddEndPuncttrue
\mciteSetBstMidEndSepPunct{\mcitedefaultmidpunct}
{\mcitedefaultendpunct}{\mcitedefaultseppunct}\relax
\EndOfBibitem
\bibitem[Peterson and Dunning(2002)Peterson, and
  Dunning]{Peterson2002_JCP_10548}
Peterson,~K.~A.; Dunning,~T.~H. {Accurate correlation consistent basis sets for
  molecular core--valence correlation effects: The second row atoms Al--Ar, and
  the first row atoms B--Ne revisited}. \emph{J. Chem. Phys.} \textbf{2002},
  \emph{117}, 10548\relax
\mciteBstWouldAddEndPuncttrue
\mciteSetBstMidEndSepPunct{\mcitedefaultmidpunct}
{\mcitedefaultendpunct}{\mcitedefaultseppunct}\relax
\EndOfBibitem
\bibitem[Prascher \latin{et~al.}(2011)Prascher, Woon, Peterson, Dunning, and
  Wilson]{Prascher2011_TCA_69}
Prascher,~B.~P.; Woon,~D.~E.; Peterson,~K.~A.; Dunning,~T.~H.; Wilson,~A.~K.
  {Gaussian basis sets for use in correlated molecular calculations. VII.
  Valence, core-valence, and scalar relativistic basis sets for Li, Be, Na, and
  Mg}. \emph{Theor. Chem. Acc.} \textbf{2011}, \emph{128}, 69--82\relax
\mciteBstWouldAddEndPuncttrue
\mciteSetBstMidEndSepPunct{\mcitedefaultmidpunct}
{\mcitedefaultendpunct}{\mcitedefaultseppunct}\relax
\EndOfBibitem
\bibitem[Ditchfield(1972)]{Ditchfield1972_JCP_5688}
Ditchfield,~R. {Molecular Orbital Theory of Magnetic Shielding and Magnetic
  Susceptibility}. \emph{J. Chem. Phys.} \textbf{1972}, \emph{56}, 5688\relax
\mciteBstWouldAddEndPuncttrue
\mciteSetBstMidEndSepPunct{\mcitedefaultmidpunct}
{\mcitedefaultendpunct}{\mcitedefaultseppunct}\relax
\EndOfBibitem
\bibitem[Becke(1988)]{Becke1988_JCP_2547}
Becke,~A.~D. A multicenter numerical integration scheme for polyatomic
  molecules. \emph{J. Chem. Phys.} \textbf{1988}, \emph{88}, 2547--2553\relax
\mciteBstWouldAddEndPuncttrue
\mciteSetBstMidEndSepPunct{\mcitedefaultmidpunct}
{\mcitedefaultendpunct}{\mcitedefaultseppunct}\relax
\EndOfBibitem
\bibitem[Treutler and Ahlrichs(1995)Treutler, and
  Ahlrichs]{Treutler1995_JCP_346}
Treutler,~O.; Ahlrichs,~R. {Efficient molecular numerical integration schemes}.
  \emph{J. Chem. Phys.} \textbf{1995}, \emph{102}, 346\relax
\mciteBstWouldAddEndPuncttrue
\mciteSetBstMidEndSepPunct{\mcitedefaultmidpunct}
{\mcitedefaultendpunct}{\mcitedefaultseppunct}\relax
\EndOfBibitem
\bibitem[Not()]{Note-1}
\Turbomole{} User's Manual,
  \url{https://www.turbomole.org/downloads/doc/Turbomole_Manual_76.pdf},
  accessed 27 October 2023.\relax
\mciteBstWouldAddEndPunctfalse
\mciteSetBstMidEndSepPunct{\mcitedefaultmidpunct}
{}{\mcitedefaultseppunct}\relax
\EndOfBibitem
\bibitem[Bahmann and Kaupp(2015)Bahmann, and Kaupp]{Bahmann2015_JCTC_1540}
Bahmann,~H.; Kaupp,~M. {Efficient Self-Consistent Implementation of Local
  Hybrid Functionals}. \emph{J. Chem. Theory Comput.} \textbf{2015}, \emph{11},
  1540--1548\relax
\mciteBstWouldAddEndPuncttrue
\mciteSetBstMidEndSepPunct{\mcitedefaultmidpunct}
{\mcitedefaultendpunct}{\mcitedefaultseppunct}\relax
\EndOfBibitem
\bibitem[Liu and Kong(2017)Liu, and Kong]{Liu2017_JCTC_2571}
Liu,~F.; Kong,~J. {Efficient Computation of Exchange Energy Density with
  Gaussian Basis Functions}. \emph{J. Chem. Theory Comput.} \textbf{2017},
  \emph{13}, 2571--2580\relax
\mciteBstWouldAddEndPuncttrue
\mciteSetBstMidEndSepPunct{\mcitedefaultmidpunct}
{\mcitedefaultendpunct}{\mcitedefaultseppunct}\relax
\EndOfBibitem
\bibitem[Laqua \latin{et~al.}(2018)Laqua, Kussmann, and
  Ochsenfeld]{Laqua2018_JCTC_3451}
Laqua,~H.; Kussmann,~J.; Ochsenfeld,~C. Efficient and Linear-Scaling
  Seminumerical Method for Local Hybrid Density Functionals. \emph{J. Chem.
  Theory Comput.} \textbf{2018}, \emph{14}, 3451--3458\relax
\mciteBstWouldAddEndPuncttrue
\mciteSetBstMidEndSepPunct{\mcitedefaultmidpunct}
{\mcitedefaultendpunct}{\mcitedefaultseppunct}\relax
\EndOfBibitem
\bibitem[Holzer(2020)]{Holzer2020_JCP_184115}
Holzer,~C. An improved seminumerical Coulomb and exchange algorithm for
  properties and excited states in modern density functional theory. \emph{J.
  Chem. Phys.} \textbf{2020}, \emph{153}, 184115\relax
\mciteBstWouldAddEndPuncttrue
\mciteSetBstMidEndSepPunct{\mcitedefaultmidpunct}
{\mcitedefaultendpunct}{\mcitedefaultseppunct}\relax
\EndOfBibitem
\bibitem[Weigend(2006)]{Weigend2006_PCCP_1057}
Weigend,~F. Accurate Coulomb-fitting basis sets for H to Rn. \emph{Phys. Chem.
  Chem. Phys.} \textbf{2006}, \emph{8}, 1057--1065\relax
\mciteBstWouldAddEndPuncttrue
\mciteSetBstMidEndSepPunct{\mcitedefaultmidpunct}
{\mcitedefaultendpunct}{\mcitedefaultseppunct}\relax
\EndOfBibitem
\bibitem[Stoychev \latin{et~al.}(2018)Stoychev, Auer, Izs{\'{a}}k, and
  Neese]{Stoychev2018_JCTC_619}
Stoychev,~G.~L.; Auer,~A.~A.; Izs{\'{a}}k,~R.; Neese,~F. Self-Consistent Field
  Calculation of Nuclear Magnetic Resonance Chemical Shielding Constants Using
  Gauge-Including Atomic Orbitals and Approximate Two-Electron Integrals.
  \emph{J. Chem. Theory Comput.} \textbf{2018}, \emph{14}, 619--637\relax
\mciteBstWouldAddEndPuncttrue
\mciteSetBstMidEndSepPunct{\mcitedefaultmidpunct}
{\mcitedefaultendpunct}{\mcitedefaultseppunct}\relax
\EndOfBibitem
\bibitem[Reiter \latin{et~al.}(2018)Reiter, Mack, and
  Weigend]{Reiter2018_JCTC_191}
Reiter,~K.; Mack,~F.; Weigend,~F. Calculation of magnetic shielding constants
  with meta-{GGA} functionals employing the multipole-accelerated resolution of
  the identity: implementation and assessment of accuracy and efficiency.
  \emph{J. Chem. Theory Comput.} \textbf{2018}, \emph{14}, 191--197\relax
\mciteBstWouldAddEndPuncttrue
\mciteSetBstMidEndSepPunct{\mcitedefaultmidpunct}
{\mcitedefaultendpunct}{\mcitedefaultseppunct}\relax
\EndOfBibitem
\bibitem[Lehtola(2021)]{Lehtola2021_JCTC_6886}
Lehtola,~S. Straightforward and Accurate Automatic Auxiliary Basis Set
  Generation for Molecular Calculations with Atomic Orbital Basis Sets.
  \emph{J. Chem. Theory Comput.} \textbf{2021}, \emph{17}, 6886--6900\relax
\mciteBstWouldAddEndPuncttrue
\mciteSetBstMidEndSepPunct{\mcitedefaultmidpunct}
{\mcitedefaultendpunct}{\mcitedefaultseppunct}\relax
\EndOfBibitem
\bibitem[Lehtola(2023)]{Lehtola2023_JCTC_6242}
Lehtola,~S. Automatic Generation of Accurate and Cost-Efficient Auxiliary Basis
  Sets. \emph{J. Chem. Theory Comput.} \textbf{2023}, \emph{19},
  6242--6254\relax
\mciteBstWouldAddEndPuncttrue
\mciteSetBstMidEndSepPunct{\mcitedefaultmidpunct}
{\mcitedefaultendpunct}{\mcitedefaultseppunct}\relax
\EndOfBibitem
\bibitem[Pedersen \latin{et~al.}(2023)Pedersen, Lehtola, Galv{\'{a}}n, and
  Lindh]{Pedersen2023_WIRCMS_1692}
Pedersen,~T.~B.; Lehtola,~S.; Galv{\'{a}}n,~I.~F.; Lindh,~R. The versatility of
  the {Cholesky} decomposition in electronic structure theory. \emph{Wiley
  Interdiscip. Rev. Comput. Mol. Sci.} \textbf{2023}, e1692\relax
\mciteBstWouldAddEndPuncttrue
\mciteSetBstMidEndSepPunct{\mcitedefaultmidpunct}
{\mcitedefaultendpunct}{\mcitedefaultseppunct}\relax
\EndOfBibitem
\bibitem[Bast(April 2020)]{Bast:20}
Bast,~R. Numgrid: Numerical integration grid for molecules. April 2020;
  \url{https://doi.org/10.5281/zenodo.1470276}\relax
\mciteBstWouldAddEndPuncttrue
\mciteSetBstMidEndSepPunct{\mcitedefaultmidpunct}
{\mcitedefaultendpunct}{\mcitedefaultseppunct}\relax
\EndOfBibitem
\bibitem[Lindh \latin{et~al.}(2001)Lindh, Malmqvist, and
  Gagliardi]{Lindh2001_TCA_178}
Lindh,~R.; Malmqvist,~P.-{\AA}.; Gagliardi,~L. {Molecular integrals by
  numerical quadrature. I. Radial integration}. \emph{Theor. Chem. Acc.}
  \textbf{2001}, \emph{106}, 178--187\relax
\mciteBstWouldAddEndPuncttrue
\mciteSetBstMidEndSepPunct{\mcitedefaultmidpunct}
{\mcitedefaultendpunct}{\mcitedefaultseppunct}\relax
\EndOfBibitem
\bibitem[Lebedev(1995)]{Lebedev1995_RASDM_283}
Lebedev,~V.~I. A quadrature formula for the sphere of 59th algebraic order of
  accuracy. \emph{Russ. Acad. Sci. Dokl. Math.} \textbf{1995}, \emph{50},
  283--286\relax
\mciteBstWouldAddEndPuncttrue
\mciteSetBstMidEndSepPunct{\mcitedefaultmidpunct}
{\mcitedefaultendpunct}{\mcitedefaultseppunct}\relax
\EndOfBibitem
\bibitem[gim()]{gimic-download}
GIMIC, version 2.0, a current density program. Can be freely downloaded from
  https://github.com/qmcurrents/gimic and
  https://zenodo.org/record/8180435\relax
\mciteBstWouldAddEndPuncttrue
\mciteSetBstMidEndSepPunct{\mcitedefaultmidpunct}
{\mcitedefaultendpunct}{\mcitedefaultseppunct}\relax
\EndOfBibitem
\bibitem[num()]{numgrid-download}
The Numgrid program. It can be freely downloaded from
  https://github.com/dftlibs/numgrid and
  https://zenodo.org/record/4815722\relax
\mciteBstWouldAddEndPuncttrue
\mciteSetBstMidEndSepPunct{\mcitedefaultmidpunct}
{\mcitedefaultendpunct}{\mcitedefaultseppunct}\relax
\EndOfBibitem
\bibitem[Arbuznikov and Kaupp(2014)Arbuznikov, and
  Kaupp]{Arbuznikov2014_JCP_204101}
Arbuznikov,~A.~V.; Kaupp,~M. {Towards improved local hybrid functionals by
  calibration of exchange-energy densities}. \emph{J. Chem. Phys.}
  \textbf{2014}, \emph{141}, 204101\relax
\mciteBstWouldAddEndPuncttrue
\mciteSetBstMidEndSepPunct{\mcitedefaultmidpunct}
{\mcitedefaultendpunct}{\mcitedefaultseppunct}\relax
\EndOfBibitem
\bibitem[Arbuznikov and Kaupp(2012)Arbuznikov, and
  Kaupp]{Arbuznikov2012_JCP_14111}
Arbuznikov,~A.~V.; Kaupp,~M. {Importance of the correlation contribution for
  local hybrid functionals: Range separation and self-interaction corrections}.
  \emph{J. Chem. Phys.} \textbf{2012}, \emph{136}, 014111\relax
\mciteBstWouldAddEndPuncttrue
\mciteSetBstMidEndSepPunct{\mcitedefaultmidpunct}
{\mcitedefaultendpunct}{\mcitedefaultseppunct}\relax
\EndOfBibitem
\bibitem[Bahmann \latin{et~al.}(2007)Bahmann, Rodenberg, Arbuznikov, and
  Kaupp]{Bahmann2007_JCP_11103}
Bahmann,~H.; Rodenberg,~A.; Arbuznikov,~A.~V.; Kaupp,~M. {A thermochemically
  competitive local hybrid functional without gradient corrections}. \emph{J.
  Chem. Phys.} \textbf{2007}, \emph{126}, 011103\relax
\mciteBstWouldAddEndPuncttrue
\mciteSetBstMidEndSepPunct{\mcitedefaultmidpunct}
{\mcitedefaultendpunct}{\mcitedefaultseppunct}\relax
\EndOfBibitem
\bibitem[Kaupp \latin{et~al.}(2007)Kaupp, Bahmann, and
  Arbuznikov]{Kaupp2007_JCP_194102}
Kaupp,~M.; Bahmann,~H.; Arbuznikov,~A.~V. Local hybrid functionals: An
  assessment for thermochemical kinetics. \emph{J. Chem. Phys.} \textbf{2007},
  \emph{127}, 194102\relax
\mciteBstWouldAddEndPuncttrue
\mciteSetBstMidEndSepPunct{\mcitedefaultmidpunct}
{\mcitedefaultendpunct}{\mcitedefaultseppunct}\relax
\EndOfBibitem
\bibitem[Arbuznikov and Kaupp(2007)Arbuznikov, and
  Kaupp]{Arbuznikov2007_CPL_160}
Arbuznikov,~A.~V.; Kaupp,~M. {Local hybrid exchange-correlation functionals
  based on the dimensionless density gradient}. \emph{Chem. Phys. Lett.}
  \textbf{2007}, \emph{440}, 160--168\relax
\mciteBstWouldAddEndPuncttrue
\mciteSetBstMidEndSepPunct{\mcitedefaultmidpunct}
{\mcitedefaultendpunct}{\mcitedefaultseppunct}\relax
\EndOfBibitem
\bibitem[Perdew \latin{et~al.}(2008)Perdew, Staroverov, Tao, and
  Scuseria]{Perdew2008_PRA_52513}
Perdew,~J.~P.; Staroverov,~V.~N.; Tao,~J.; Scuseria,~G.~E. {Density functional
  with full exact exchange, balanced nonlocality of correlation, and constraint
  satisfaction}. \emph{Phys. Rev. A} \textbf{2008}, \emph{78}, 052513\relax
\mciteBstWouldAddEndPuncttrue
\mciteSetBstMidEndSepPunct{\mcitedefaultmidpunct}
{\mcitedefaultendpunct}{\mcitedefaultseppunct}\relax
\EndOfBibitem
\bibitem[Johnson(2014)]{Johnson2014_JCP_124120}
Johnson,~E.~R. Local-hybrid functional based on the correlation length.
  \emph{J. Chem. Phys.} \textbf{2014}, \emph{141}, 124120\relax
\mciteBstWouldAddEndPuncttrue
\mciteSetBstMidEndSepPunct{\mcitedefaultmidpunct}
{\mcitedefaultendpunct}{\mcitedefaultseppunct}\relax
\EndOfBibitem
\bibitem[Mardirossian and Head-Gordon(2015)Mardirossian, and
  Head-Gordon]{Mardirossian2015_074111}
Mardirossian,~N.; Head-Gordon,~M. Mapping the genome of meta-generalized
  gradient approximation density functionals: The search for {B97M}-{V}.
  \emph{J. Chem. Phys.} \textbf{2015}, \emph{142}, 074111\relax
\mciteBstWouldAddEndPuncttrue
\mciteSetBstMidEndSepPunct{\mcitedefaultmidpunct}
{\mcitedefaultendpunct}{\mcitedefaultseppunct}\relax
\EndOfBibitem
\bibitem[{Van Voorhis} and Scuseria(1998){Van Voorhis}, and
  Scuseria]{VanVoorhis1998_JCP_400}
{Van Voorhis},~T.; Scuseria,~G.~E. {A novel form for the exchange-correlation
  energy functional}. \emph{J. Chem. Phys.} \textbf{1998}, \emph{109},
  400\relax
\mciteBstWouldAddEndPuncttrue
\mciteSetBstMidEndSepPunct{\mcitedefaultmidpunct}
{\mcitedefaultendpunct}{\mcitedefaultseppunct}\relax
\EndOfBibitem
\bibitem[Yu \latin{et~al.}(2016)Yu, He, and Truhlar]{Yu2016_1280}
Yu,~H.~S.; He,~X.; Truhlar,~D.~G. {MN15}-{L}: A New Local Exchange-Correlation
  Functional for {Kohn}--{Sham} Density Functional Theory with Broad Accuracy
  for Atoms, Molecules, and Solids. \emph{J. Chem. Theory Comput.}
  \textbf{2016}, \emph{12}, 1280--1293\relax
\mciteBstWouldAddEndPuncttrue
\mciteSetBstMidEndSepPunct{\mcitedefaultmidpunct}
{\mcitedefaultendpunct}{\mcitedefaultseppunct}\relax
\EndOfBibitem
\bibitem[Tao \latin{et~al.}(2003)Tao, Perdew, Staroverov, and
  Scuseria]{Tao2003_146401}
Tao,~J.; Perdew,~J.~P.; Staroverov,~V.~N.; Scuseria,~G.~E. Climbing the Density
  Functional Ladder: Nonempirical Meta-Generalized Gradient Approximation
  Designed for Molecules and Solids. \emph{Phys. Rev. Lett.} \textbf{2003},
  \emph{91}, 146401\relax
\mciteBstWouldAddEndPuncttrue
\mciteSetBstMidEndSepPunct{\mcitedefaultmidpunct}
{\mcitedefaultendpunct}{\mcitedefaultseppunct}\relax
\EndOfBibitem
\bibitem[Perdew \latin{et~al.}(2004)Perdew, Tao, Staroverov, and
  Scuseria]{Perdew2004_6898}
Perdew,~J.~P.; Tao,~J.; Staroverov,~V.~N.; Scuseria,~G.~E. Meta-generalized
  gradient approximation: Explanation of a realistic nonempirical density
  functional. \emph{J. Chem. Phys.} \textbf{2004}, \emph{120}, 6898\relax
\mciteBstWouldAddEndPuncttrue
\mciteSetBstMidEndSepPunct{\mcitedefaultmidpunct}
{\mcitedefaultendpunct}{\mcitedefaultseppunct}\relax
\EndOfBibitem
\bibitem[Zhao and Truhlar(2006)Zhao, and Truhlar]{Zhao2006_194101}
Zhao,~Y.; Truhlar,~D.~G. A new local density functional for main-group
  thermochemistry, transition metal bonding, thermochemical kinetics, and
  noncovalent interactions. \emph{J. Chem. Phys.} \textbf{2006}, \emph{125},
  194101\relax
\mciteBstWouldAddEndPuncttrue
\mciteSetBstMidEndSepPunct{\mcitedefaultmidpunct}
{\mcitedefaultendpunct}{\mcitedefaultseppunct}\relax
\EndOfBibitem
\bibitem[Boese and Handy(2002)Boese, and Handy]{Boese2002_JCP_9559}
Boese,~A.~D.; Handy,~N.~C. {New exchange-correlation density functionals: The
  role of the kinetic-energy density}. \emph{J. Chem. Phys.} \textbf{2002},
  \emph{116}, 9559--9569\relax
\mciteBstWouldAddEndPuncttrue
\mciteSetBstMidEndSepPunct{\mcitedefaultmidpunct}
{\mcitedefaultendpunct}{\mcitedefaultseppunct}\relax
\EndOfBibitem
\bibitem[Zhao and Truhlar(2005)Zhao, and Truhlar]{Zhao2005_5656}
Zhao,~Y.; Truhlar,~D.~G. Design of Density Functionals That Are Broadly
  Accurate for Thermochemistry, Thermochemical Kinetics, and Nonbonded
  Interactions. \emph{J. Phys. Chem. A} \textbf{2005}, \emph{109}, 5656\relax
\mciteBstWouldAddEndPuncttrue
\mciteSetBstMidEndSepPunct{\mcitedefaultmidpunct}
{\mcitedefaultendpunct}{\mcitedefaultseppunct}\relax
\EndOfBibitem
\bibitem[Zhao and Truhlar(2008)Zhao, and Truhlar]{Zhao2008_215}
Zhao,~Y.; Truhlar,~D.~G. The {M06} suite of density functionals for main group
  thermochemistry, thermochemical kinetics, noncovalent interactions, excited
  states, and transition elements: two new functionals and systematic testing
  of four {M06}-class functionals and 12 other functionals. \emph{Theor. Chem.
  Acc.} \textbf{2008}, \emph{120}, 215\relax
\mciteBstWouldAddEndPuncttrue
\mciteSetBstMidEndSepPunct{\mcitedefaultmidpunct}
{\mcitedefaultendpunct}{\mcitedefaultseppunct}\relax
\EndOfBibitem
\bibitem[Yu \latin{et~al.}(2016)Yu, He, Li, and Truhlar]{Yu2016_5032}
Yu,~H.~S.; He,~X.; Li,~S.~L.; Truhlar,~D.~G. {MN15}: A {Kohn}--{Sham}
  global-hybrid exchange-correlation density functional with broad accuracy for
  multi-reference and single-reference systems and noncovalent interactions.
  \emph{Chem. Sci.} \textbf{2016}, \emph{7}, 5032--5051\relax
\mciteBstWouldAddEndPuncttrue
\mciteSetBstMidEndSepPunct{\mcitedefaultmidpunct}
{\mcitedefaultendpunct}{\mcitedefaultseppunct}\relax
\EndOfBibitem
\bibitem[Stephens \latin{et~al.}(1994)Stephens, Devlin, Chabalowski, and
  Frisch]{Stephens1994_11623}
Stephens,~P.~J.; Devlin,~F.~J.; Chabalowski,~C.~F.; Frisch,~M.~J. Ab Initio
  Calculation of Vibrational Absorption and Circular Dichroism Spectra Using
  Density Functional Force Fields. \emph{J. Phys. Chem.} \textbf{1994},
  \emph{98}, 11623\relax
\mciteBstWouldAddEndPuncttrue
\mciteSetBstMidEndSepPunct{\mcitedefaultmidpunct}
{\mcitedefaultendpunct}{\mcitedefaultseppunct}\relax
\EndOfBibitem
\bibitem[Hertwig and Koch(1997)Hertwig, and Koch]{Hertwig1997_CPL_345}
Hertwig,~R.~H.; Koch,~W. On the parameterization of the local correlation
  functional. What is {Becke}-3-{LYP}? \emph{Chem. Phys. Lett.} \textbf{1997},
  \emph{268}, 345--351\relax
\mciteBstWouldAddEndPuncttrue
\mciteSetBstMidEndSepPunct{\mcitedefaultmidpunct}
{\mcitedefaultendpunct}{\mcitedefaultseppunct}\relax
\EndOfBibitem
\bibitem[Becke(1993)]{Becke1993_1372}
Becke,~A.~D. A new mixing of {Hartree}--{Fock} and local density-functional
  theories. \emph{J. Chem. Phys.} \textbf{1993}, \emph{98}, 1372\relax
\mciteBstWouldAddEndPuncttrue
\mciteSetBstMidEndSepPunct{\mcitedefaultmidpunct}
{\mcitedefaultendpunct}{\mcitedefaultseppunct}\relax
\EndOfBibitem
\bibitem[Mardirossian and Head-Gordon(2016)Mardirossian, and
  Head-Gordon]{Mardirossian2016_JCP_214110}
Mardirossian,~N.; Head-Gordon,~M. $\omega${B97M-V}: A combinatorially
  optimized, range-separated hybrid, meta-{GGA} density functional with {VV10}
  nonlocal correlation. \emph{J. Chem. Phys.} \textbf{2016}, \emph{144},
  214110\relax
\mciteBstWouldAddEndPuncttrue
\mciteSetBstMidEndSepPunct{\mcitedefaultmidpunct}
{\mcitedefaultendpunct}{\mcitedefaultseppunct}\relax
\EndOfBibitem
\bibitem[Mardirossian and Head-Gordon(2014)Mardirossian, and
  Head-Gordon]{Mardirossian2014_9904}
Mardirossian,~N.; Head-Gordon,~M. $\omega${B97X}-{V}: A 10-parameter,
  range-separated hybrid, generalized gradient approximation density functional
  with nonlocal correlation, designed by a survival-of-the-fittest strategy.
  \emph{Phys. Chem. Chem. Phys.} \textbf{2014}, \emph{16}, 9904--9924\relax
\mciteBstWouldAddEndPuncttrue
\mciteSetBstMidEndSepPunct{\mcitedefaultmidpunct}
{\mcitedefaultendpunct}{\mcitedefaultseppunct}\relax
\EndOfBibitem
\bibitem[Lehtola(2023)]{Lehtola2023_JCTC_2502}
Lehtola,~S. Meta-{GGA} Density Functional Calculations on Atoms with
  Spherically Symmetric Densities in the Finite Element Formalism. \emph{J.
  Chem. Theory Comput.} \textbf{2023}, \emph{19}, 2502--2517\relax
\mciteBstWouldAddEndPuncttrue
\mciteSetBstMidEndSepPunct{\mcitedefaultmidpunct}
{\mcitedefaultendpunct}{\mcitedefaultseppunct}\relax
\EndOfBibitem
\bibitem[Lehtola(2023)]{Lehtola2023_JPCA_4180}
Lehtola,~S. Atomic Electronic Structure Calculations with {Hermite}
  Interpolating Polynomials. \emph{J. Phys. Chem. A} \textbf{2023}, \emph{127},
  4180--4193\relax
\mciteBstWouldAddEndPuncttrue
\mciteSetBstMidEndSepPunct{\mcitedefaultmidpunct}
{\mcitedefaultendpunct}{\mcitedefaultseppunct}\relax
\EndOfBibitem
\bibitem[Ruud \latin{et~al.}(1997)Ruud, Helgaker, and
  J{\o}rgensen]{Ruud1997_JCP_10599}
Ruud,~K.; Helgaker,~T.; J{\o}rgensen,~P. {The effect of correlation on
  molecular magnetizabilities and rotational g tensors}. \emph{J. Chem. Phys.}
  \textbf{1997}, \emph{107}, 10599\relax
\mciteBstWouldAddEndPuncttrue
\mciteSetBstMidEndSepPunct{\mcitedefaultmidpunct}
{\mcitedefaultendpunct}{\mcitedefaultseppunct}\relax
\EndOfBibitem
\bibitem[van Wüllen(1992)]{vanWuellenThesis}
van Wüllen,~C. Die Berechnung magnetischer Eigenschaften unter
  Berücksichtigung der Elektronenkorrelation: Die
  Multikonfigurations-Verallgemeinerung der IGLO Methode. Dissertation,
  Ruhr-Universität Bochum, 1992\relax
\mciteBstWouldAddEndPuncttrue
\mciteSetBstMidEndSepPunct{\mcitedefaultmidpunct}
{\mcitedefaultendpunct}{\mcitedefaultseppunct}\relax
\EndOfBibitem
\bibitem[Janesko \latin{et~al.}(2017)Janesko, Proynov, Kong, Scalmani, and
  Frisch]{Janesko2017_JPCL_4314}
Janesko,~B.~G.; Proynov,~E.; Kong,~J.; Scalmani,~G.; Frisch,~M.~J. Practical
  Density Functionals beyond the Overdelocalization–Underbinding Zero-Sum
  Game. \emph{J. Phys. Chem. Lett.} \textbf{2017}, \emph{8}, 4314--4318\relax
\mciteBstWouldAddEndPuncttrue
\mciteSetBstMidEndSepPunct{\mcitedefaultmidpunct}
{\mcitedefaultendpunct}{\mcitedefaultseppunct}\relax
\EndOfBibitem
\end{mcitethebibliography}
\end{document}